\pdfoutput=1
\documentclass[11pt,twoside,a4paper,cmspaper,final,collab]{cms-tdr}
\def\svnVersion{14ecfe3}\def\svnDate{2022/04/18}\def\cmsCernNoTag{CERN-EP-2021-266}\def\cmsCernDate{\today}\def\cmsMessage{Published in the Journal of High Energy Physics as \href{http://dx.doi.org/10.1007/JHEP04(2022)062}{\doi{10.1007/JHEP04(2022)062}.}}
\begin{document}\cmsNoteHeader{EXO-20-014}

\newcommand{\JHUGen} {{\textsc{jhugen}}\xspace}
\newcommand{\lumivalscouting}{101\fbinv}
\newcommand{\lumivalseventeen}{41.5\fbinv}
\newcommand{\lumivaleighteen}{59.7\fbinv}
\newcommand{\mmumu}{\ensuremath{m_{\PGm\PGm}}\xspace}
\newcommand{\ptmu}{\ensuremath{\pt^{\PGm}}\xspace}
\newcommand{\etamu}{\ensuremath{\eta^{\PGm}}\xspace}
\newcommand{\ptmuone}{\ensuremath{\pt^{\PGm_{\mathrm{1}}}}\xspace}
\newcommand{\ptmutwo}{\ensuremath{\pt^{\PGm_{\mathrm{2}}}}\xspace}
\newcommand{\ptmumu}{\ensuremath{\pt^{\PGm\PGm}}\xspace}
\newcommand{\ptmumuvec}{\ensuremath{\vec{p}_{\, \mathrm{T}}^{\PGm\PGm}}\xspace}
\newcommand{\drmumu}{\ensuremath{\Delta R(\PGm_{\mathrm{1}},\PGm_{\mathrm{2}})}\xspace}
\newcommand{\dphimumu}{\ensuremath{\Delta\phi(\PGm_{\mathrm{1}},\PGm_{\mathrm{2}})}\xspace}
\newcommand{\dphimudv}{\ensuremath{\Delta\phi(\ptmumuvec,\overrightarrow{\mathrm{SV}})}\xspace}
\newcommand{\vecdv}{\ensuremath{\overrightarrow{\mathrm{SV}}}\xspace}
\newcommand{\detamumu}{\ensuremath{\Delta\eta(\PGm_{\mathrm{1}},\PGm_{\mathrm{2}})}\xspace}
\newcommand{\logabsetaphi}{\ensuremath{\log_{10}(|\Delta\eta_{\mu\mu}|/|\Delta\phi_{\mu\mu}|)}\xspace}
\newcommand{\lxy}{\ensuremath{l_\mathrm{xy}}\xspace}
\newcommand{\dxy}{\ensuremath{d_\mathrm{xy}}\xspace}
\newcommand{\sigmadxy}{\ensuremath{\sigma_{\dxy}}\xspace}
\newcommand{\scaleddxy}{\ensuremath{\abs{\dxy}/(\lxy\mmumu/\ptmumu)}\xspace}

\cmsNoteHeader{EXO-20-014}
\title{Search for long-lived particles decaying into muon pairs in proton-proton collisions at \texorpdfstring{$\sqrt{s} = 13\TeV$}{sqrt(s) = 13 TeV} collected with a dedicated high-rate data stream}

\date{\today}

\abstract{
A search for long-lived particles decaying into muon pairs is performed using proton-proton collisions at a center-of-mass energy of 13\TeV, collected by the CMS experiment at the LHC in 2017 and 2018, corresponding to an integrated luminosity of \lumivalscouting. The data sets used in this search were collected with a dedicated dimuon trigger stream with low transverse momentum thresholds, recorded at high rate by retaining a reduced amount of information, in order to explore otherwise inaccessible phase space at low dimuon mass and nonzero displacement from the primary interaction vertex. No significant excess of events beyond the standard model expectation is found. Upper limits on branching fractions at 95\% confidence level are set on a wide range of mass and lifetime hypotheses in beyond the standard model frameworks with the Higgs boson decaying into a pair of long-lived dark photons, or with a long-lived scalar resonance arising from a decay of a \PQb hadron. The limits are the most stringent to date for substantial regions of the parameter space. These results can be also used to constrain models of displaced dimuons that are not explicitly considered in this paper.
}

\hypersetup{%
pdfauthor={CMS Collaboration},%
pdftitle={Search for long-lived particles decaying into muon pairs in proton-proton collisions at sqrt(s) = 13 TeV collected with a dedicated high-rate data stream},%
pdfsubject={CMS},%
pdfkeywords={CMS,  exotica, long-lived, LLP, low-mass, resonance}}

\maketitle

\section{Introduction}
\label{sec:introduction}

Cosmological evidence points to the existence of dark matter~\cite{DM1,DM2,DM3,DM4}, whose origin remains one of the outstanding problems in particle physics and cosmology.
Beyond gravitational interactions, dark matter is expected to interact very weakly, if at all, with standard model (SM) particles.
This introduces the possibility of a hidden (dark) sector of matter~\cite{DM5,DM6}.
Particles in the dark sector would interact with the SM particles only via weakly interacting mediators.

\begin{figure}[ht]
  \centering 
    \vspace{0.01\textwidth}
    \includegraphics[width=0.4\textwidth]{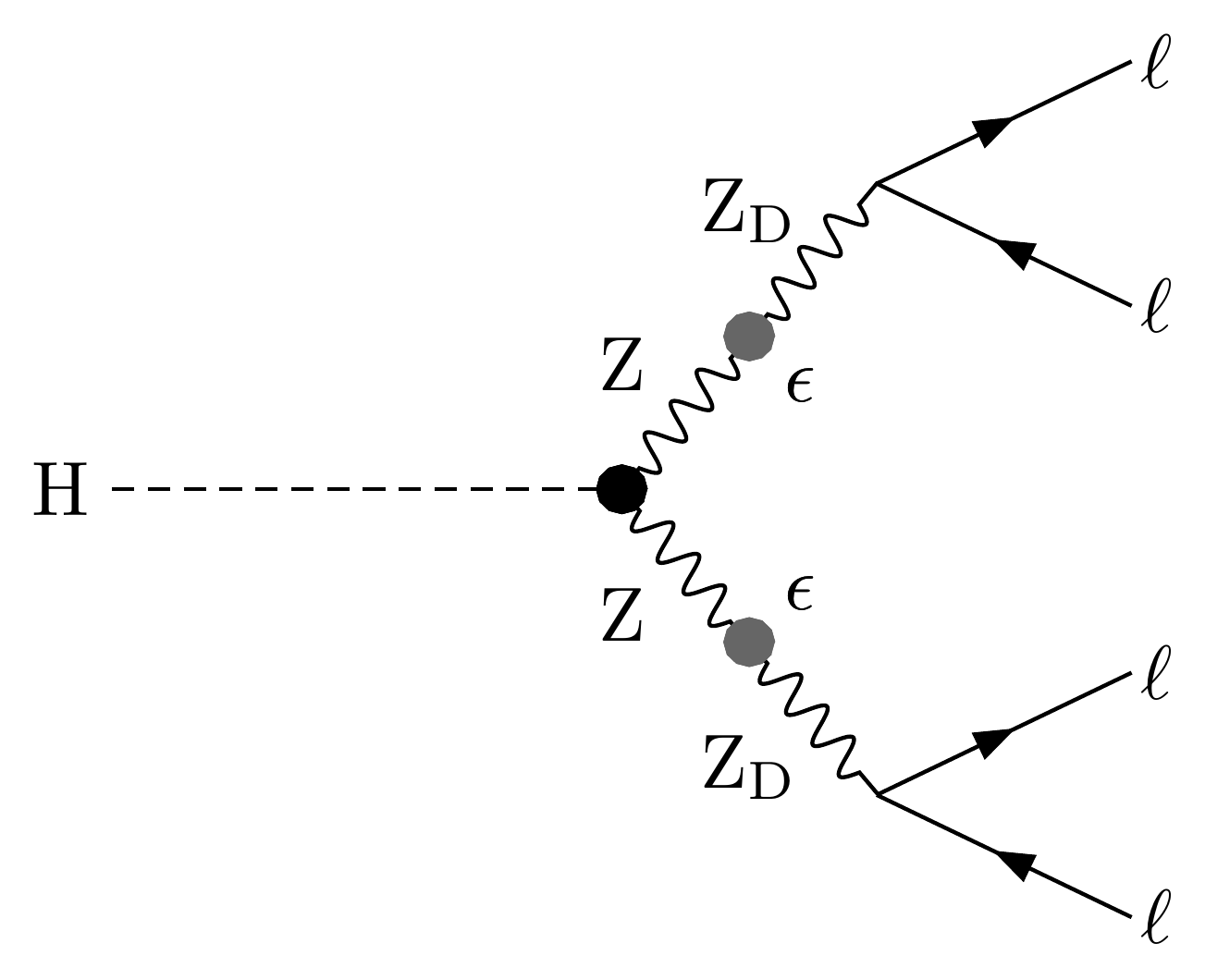}
    \hspace{0.1\textwidth}
    \includegraphics[width=0.4\textwidth]{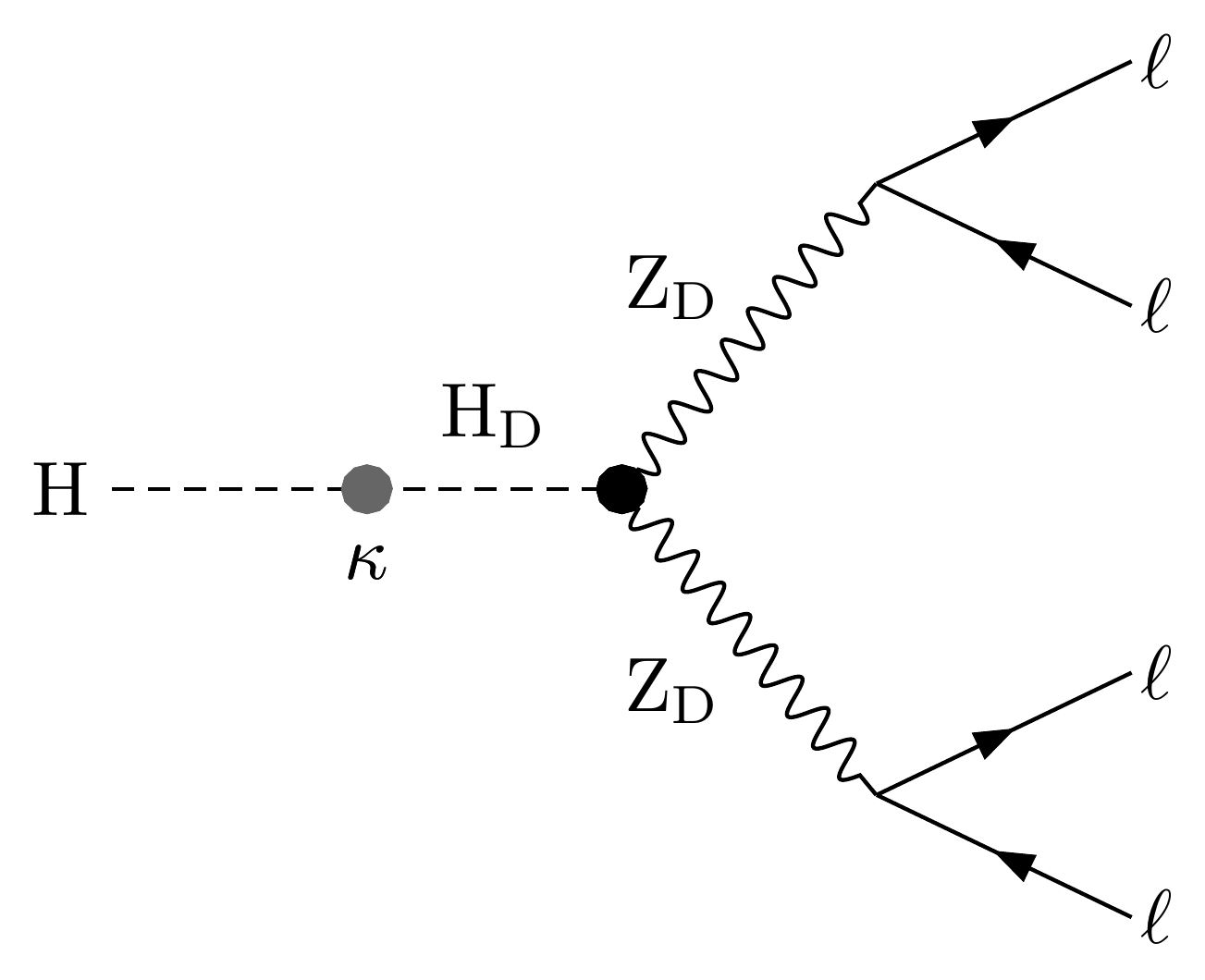}
    \vspace{0.01\textwidth}
    \caption{Diagrams illustrating an SM-like Higgs boson (\PH) decay to four leptons ($\ell$) 
      via two intermediate dark photons, $\PZ_{\mathrm{D}}$~\cite{DM6}:
      (left) through the hypercharge portal; (right) through the Higgs portal, via a dark Higgs boson ($\PH_{\mathrm{D}}$). 
      \label{fig:HZdZd}} 
\end{figure}

One dark matter scenario involves a spontaneously broken dark $U(1)_{\mathrm{D}}$ 
gauge symmetry, 
mediated by a dark photon, $\PZ_{\mathrm{D}}$~\cite{DM6}.
In this scenario, the only renormalizable interaction with SM particles is through kinetic mixing of the dark photon with the hypercharge gauge boson.
In addition, if a dark Higgs mechanism is responsible for the spontaneous breaking of the $U(1)_{\mathrm{D}}$ gauge symmetry, 
then the dark Higgs boson ($\PH_{\mathrm{D}}$) has a renormalizable coupling to the 125\GeV SM-like Higgs boson (\PH), 
resulting in mixing between the two physical scalar states.
Thus, the hidden sector may interact with the SM either through the hypercharge portal via the kinetic mixing coupling (denoted as $\epsilon$),
or through the Higgs portal via the Higgs mixing (denoted as $\kappa$).
The dark photon $\PZ_{\mathrm{D}}$ may also mix with the SM photon (\PGg) and the \PZ boson through the hypercharge portal.
In the absence of hidden-sector states below the $\PZ_{\mathrm{D}}$ mass, this mixing causes the $\PZ_{\mathrm{D}}$ to decay exclusively to SM particles, 
with a sizable branching fraction to leptons, with the coupling of the SM fermions to $\PZ_{\mathrm{D}}$ being proportional to $\epsilon$.
The $\PZ_{\mathrm{D}}$ boson is expected to be long-lived if $\epsilon\lesssim10^{-4}$. 
In this range of parameter values, the decays $\PH\to\PZ\PZ_{\mathrm{D}}$ and $\PH\to\PZ_{\mathrm{D}}\PZ_{\mathrm{D}}$ 
through the hypercharge portal have negligible branching fractions.
Diagrams in Fig.~\ref{fig:HZdZd} illustrate the production of two $\PZ_{\mathrm{D}}$ bosons from a Higgs boson.
Constraints have been placed on the visible dark photon decays by beam dump~\cite{ZDLimits-1},
fixed-target~\cite{ZDLimits-2}, $\Pep\Pem$ collider~\cite{ZDLimits-3}, and rare-decay experiments~\cite{ZDLimits-4}, 
as well as by the LHCb~\cite{ZDLimits-5a,ZDLimits-5b,ZDLimits-5c} and CMS~\cite{EXO-19-018} experiments at the CERN LHC. 

\begin{figure}[ht]
  \centering 
    \includegraphics[width=0.4\textwidth]{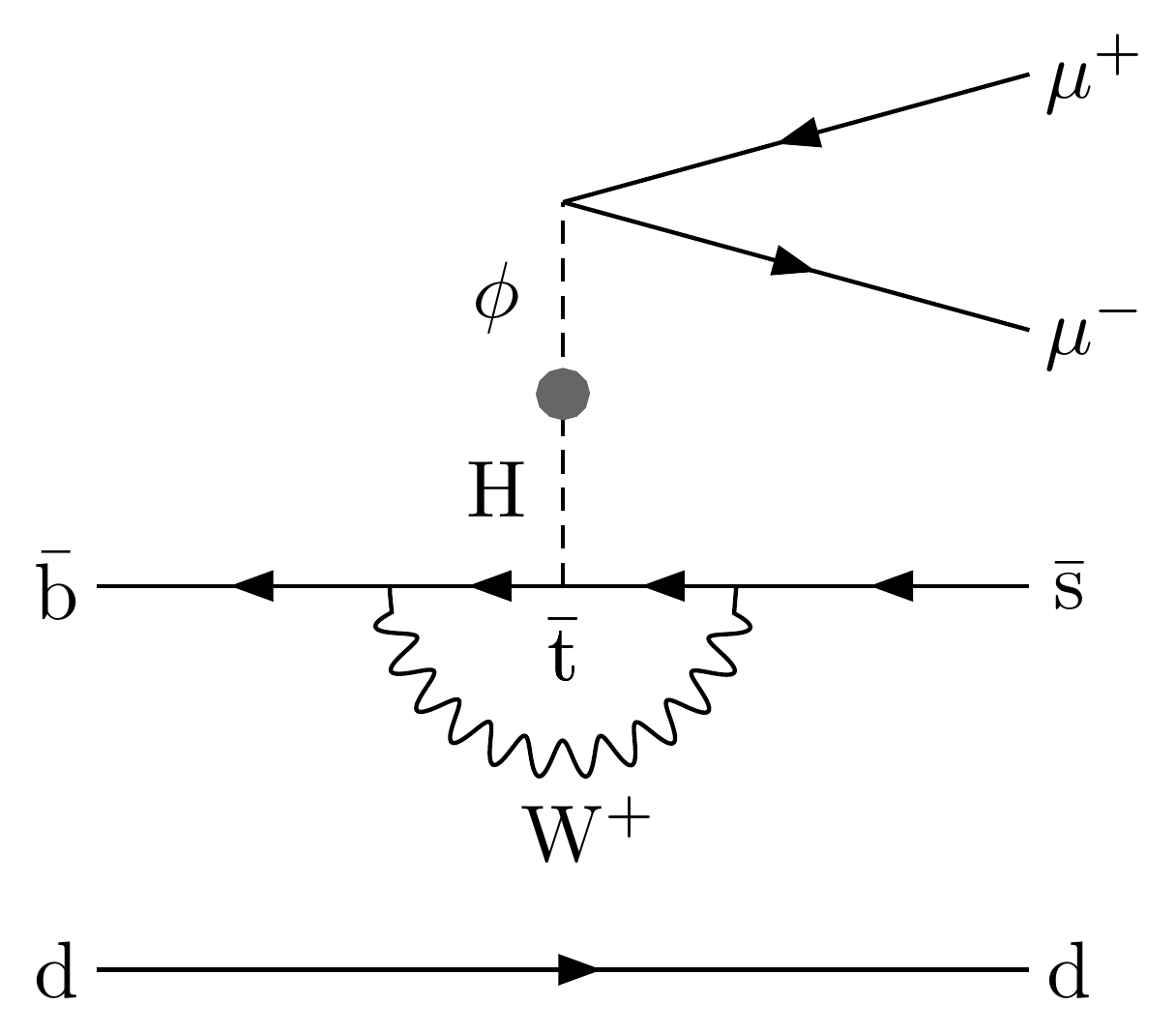}
    \caption{Diagram illustrating the production of a scalar resonance $\phi$ 
      in a \PQb hadron decay, through mixing with an SM-like Higgs boson (\PH).
      \label{fig:PhiFeyn}}
\end{figure}

Other scenarios may produce a low-mass, long-lived resonance decaying into a muon pair.
For instance, one of the minimal extensions to the SM adds a singlet scalar field $\phi$, 
which mixes with the \PH boson and couples to all SM fermions~\cite{BPhi-inflaton,BPhi}.
In the hypothesis of weak coupling to SM fermions with a signal mixing angle $\theta\lesssim10^{-3}$, $\phi$ will be long-lived.
Such a scalar resonance may be produced in the decay of a \PQb hadron, $h_{\PQb} \to \phi X$, 
as illustrated in Fig.~\ref{fig:PhiFeyn}.
Constraints on this model have been previously placed by the CHARM~\cite{BtoPhiLimits-0} and LHCb~\cite{BtoPhiLimits-1,BtoPhiLimits-2} experiments.

This paper presents a search for narrow, long-lived dimuon resonances based on
dimuon data collected with
the CMS experiment during the CERN LHC Run 2 in 2017 and 2018
using a dedicated dimuon trigger stream (\textit{scouting}) with low transverse momentum (\pt) thresholds, 
recorded at high rate by retaining a reduced amount of information.
The rate of scouting triggers is higher than that 
of the standard triggers owing to less stringent requirements 
that enable dimuon resonance searches across mass and lifetime ranges 
that are otherwise inaccessible.
The selected data correspond to a total integrated luminosity of \lumivalscouting (\lumivalseventeen collected in 2017 and \lumivaleighteen collected in 2018).
Additional details on the data sets and triggers are provided in Sections~\ref{sec:cmsdetector} and~\ref{sec:triggers}.
The search targets narrow, low-mass, long-lived resonances decaying into a pair of oppositely charged muons, 
where the lifetime of the long-lived particle is such that the transverse displacement (\lxy) of its decay vertex is within 11\cm 
of the primary interaction vertex (PV).
The signal is expected to appear as a narrow peak on the dimuon mass continuum, 
with a resonance width smaller than the experimental mass resolution. 

\section{The CMS detector}
\label{sec:cmsdetector}

The central feature of the CMS apparatus is a superconducting solenoid of 6\unit{m} internal diameter
providing a magnetic field of 3.8\unit{T}. Within the solenoid volume are a silicon pixel and strip tracker, a
lead tungstate crystal electromagnetic calorimeter, and a brass and scintillator hadron calorimeter, each composed of
a barrel and two endcap sections. Forward calorimeters extend the pseudorapidity ($\eta$) coverage provided by the
 barrel and endcap detectors. Muons are measured in gas-ionization detectors embedded in the steel flux-return yoke outside
the solenoid. 
The pixel tracker consists of four concentric barrel layers at radii of 29, 68, 109, and 160\mm, 
and three disks on each end at distances of 291, 396, and 516\mm from the center of the detector~\cite{phase1tracker}.
A more detailed description of the CMS detector, together with a definition of the coordinate system used and the relevant
kinematic variables, can be found in Ref.~\cite{Chatrchyan:2008zzk}.

A two-level trigger system is used to select events of potential physics interest.
 The first level of the CMS trigger system (L1)~\cite{CMSL1}, composed of custom hardware processors, uses information from the
calorimeters and muon detectors to select events of interest in a time interval of less than 4\mus. The high-level
 trigger (HLT) processor farm further decreases the event rate from around 100\unit{kHz} to about 1\unit{kHz}, before data storage.
A more detailed description of the CMS trigger system can be found in Ref.~\cite{CMStrigger}.

In addition to the standard HLT selection streams, 
a dedicated set of triggers exists that allows the exploration of
otherwise inaccessible phase space. 
This approach is referred to as \textit{data scouting}~\cite{EXO-14-005,EXO-19-018}.
The scouting trigger algorithms used in this search
select events containing muon pairs with 
a mass $\mmumu\gtrsim 2 m_{\PGm}$ at a rate of about 3\unit{kHz}.
In order to compensate for the large rate,
the event size is reduced by up to a factor of 1000
by only retaining limited information, as reconstructed at the HLT. 

\section{Data sets and triggers}
\label{sec:triggers}

The standard trigger streams typically select events containing at least one particle (\eg, a lepton) or jet with large \pt.
Thus, such triggers are not necessarily ideal in a generic search for light narrow resonances decaying into two muons, 
with $\mmumu\lesssim45\GeV$~\cite{EXO-19-018}.
The scouting triggers used in this search compensate for the large
trigger rate by recording only limited event information
reconstructed at the HLT.
For muons, 
which are reconstructed by combining the information from the tracker and the muon detectors,
the recorded information includes for each muon its four-momentum, 
the number of hits left by each muon track in the tracker, 
the normalized $\chi^2$ of the muon track, and
the isolation of the muon, 
computed as the scalar sum of the \pt 
of all additional tracks in a cone of size 
$\Delta R = \sqrt{\smash[b]{(\Delta\eta)^2+(\Delta\phi)^2}} = 0.3$ around the muon, 
where $\phi$ is the azimuthal angle with respect to the counterclockwise beam axis. 
Muon tracks are required to leave a hit in at least two layers or disks of the pixel tracker
and are used in pairs to form dimuon vertices, considering all possible pairs.
In the following, these vertices are 
termed secondary vertices (SVs),
and they may be displaced from the PV or not.
Candidate proton-proton ($\Pp\Pp$) interaction vertices are reconstructed 
using any track with a hit in each layer or disk of the pixel tracker, 
and the candidate vertex with the largest value of summed track $\pt^{2}$ is taken to be
the PV of the $\Pp\Pp$ interaction.
Events in the scouting data sets contain information on 
the normalized $\chi^2$ of all reconstructed vertices, 
together with their position with respect to the nominal interaction point.
The four-momenta of jets formed by clustering calorimeter deposits 
using the anti-\kt algorithm~\cite{Cacciari:2008gp, Cacciari:2011ma} 
with a distance parameter of 0.4 are also available, 
for jets with $\pt>20\GeV$ and $\abs{\eta}<3.0$.

Events are preselected at the L1 trigger by requiring data collected in 2017 and 2018 to pass at least one of three criteria:
\begin{itemize}
\item the presence of two L1 muons with opposite charges (OS),
$\ptmu>4\GeV$ (for 2017 data) or 4.5\GeV (for 2018 data) and $\drmumu<1.2$;
\item the presence of two OS L1 muons with  $\abs{\etamu}<1.4$ and $\drmumu<1.4$;
\item the presence of two L1 muons with $\ptmuone>15$ and $\ptmutwo>7\GeV$.
\end{itemize}
Then, only the preselected events containing two OS muons 
with $\ptmu>3\GeV$ and $\abs{\etamu}<2.4$ at the HLT are retained.

The efficiency of the scouting triggers is measured in data, as a function 
of the dimuon \lxy and 
of the minimum \ptmu of the pair,
using events with at least two muons selected with independent standard triggers.
The measured efficiency and its uncertainty
are accounted for in the normalization of simulated events, as discussed in Section~\ref{sec:mc}.

The resolution in \ptmu for muons reconstructed at the HLT is measured to be roughly 10\% worse relative to muons undergoing offline reconstruction, 
because of differences in the calibration of the HLT and the offline reconstruction.
As only a reduced amount of information is available in scouting data, 
no additional calibration is performed for the muons used in this search.

\section{Monte Carlo simulation}
\label{sec:mc}

The SM background is estimated directly from data as a continuum background 
in the \mmumu spectrum parametrized by analytical functions.
Thus, the search does not rely on Monte Carlo (MC) simulation 
in order to estimate the background.

On the other hand, MC simulations of beyond the SM (BSM) signals are used to interpret the results of the search.
Two signal models are simulated, as described in Section~\ref{sec:introduction}:
\begin{itemize}
\item A model of \PH boson production via the dominant mode of gluon-gluon fusion, 
with $\PH\to\PZ_{\mathrm{D}}\PZ_{\mathrm{D}}$,
where $\PZ_{\mathrm{D}}$ is a long-lived resonance and 
at least one $\PZ_{\mathrm{D}}$ decays to a pair of oppositely charged muons.
This model is generated   
using the \POWHEG~$\mathrm{v}2.0$~\cite{powheg0,powheg1,powheg2} generator at next-to-leading order (NLO) precision 
and the \JHUGen~7.4.0~\cite{jhugen1,jhugen2,jhugen3,jhugen4,jhugen5} generator at leading order (LO) precision.
The \JHUGen generator is used to model the \PH boson decay, setting the properties (mass and lifetime) of the $\PZ_{\mathrm{D}}$ boson.
The generators are interfaced with \PYTHIA~8.230~\cite{pythia82} 
for the simulation of parton showering and hadronization.
The simulated signal is normalized to a Higgs production cross section calculated at next-to-next-to-NLO (N$^3$LO) 
in the strong coupling \alpS~\cite{HiggsXS13TeV}.
The decays $\PH\to\PZ\PZ_{\mathrm{D}}$ and $\PH\to\PZ_{\mathrm{D}}\PZ_{\mathrm{D}}$ through the hypercharge portal 
have negligible branching fractions for parameter values explored in this analysis.
\item A model in which a scalar resonance $\phi$ is emitted in the decay of \PQb hadrons, 
where $\phi$ is long lived and decays to a pair of oppositely charged muons with 100\% probability. 
The production of \PQb hadrons is initially generated at LO precision using \PYTHIA~8.2.
Subsequently, a fixed-order calculation at next-to-leading log (FONLL) 
in \alpS~\cite{FONLL_1,FONLL_2,FONLL_3,FONLL_4, Khachatryan_2017}
is used to reweight the \pt spectrum of \PQb hadrons decaying into $\phi$.
The simulated signal is also normalized to the FONLL calculation.
\end{itemize}

The generators use the CP5 \PYTHIA tune~\cite{CP15bib} and the
NNPDF3.1 sets of parton distribution functions (PDFs)~\cite{nnpdf3bib}, while
the detector response is simulated with a \GEANTfour model~\cite{Geant} 
of the CMS detector.
The products from multiple $\Pp\Pp$~interactions are superimposed on those from the hard collision (pileup),
and the simulated samples are reweighted such that 
the number of collisions per bunch crossing reflects
the distribution observed in data.

The search explores a dimuon mass range from about 200\MeV (\ie, $\mmumu\gtrsim 2 m_{\PGm}$) to about 50\GeV,
and a transverse displacement range $0<\lxy<11\cm$ from the PV.
The hard cutoff in \lxy at 11\cm is due to the definition of the scouting trigger stream,
which requires the presence of hits in at least two layers of the pixel tracker.
Signal samples are generated to cover the full ranges of interest, in both
the \mmumu and the $c\tau_{0}$ (mean proper decay length) dimensions.
Only masses $m_{\phi}\lesssim5\GeV$ can be probed
in the model where a scalar resonance ($\phi$) is produced on shell in the decay of a \PQb hadron.

\section{Event selection and categorization}
\label{sec:selection}

\subsection{Event selection}
\label{sec:evtsel}

The events collected using the scouting triggers (see Section~\ref{sec:triggers})
are further selected to contain at least one pair of OS muons
associated with an SV
that satisfies a number of quality criteria:
\begin{itemize}
\item $\lxy < 11\unit{cm}$;
\item $x$ coordinate uncertainty $<0.05\unit{cm}$;
\item $y$ coordinate uncertainty $<0.05\unit{cm}$;
\item $z$ coordinate uncertainty $<0.10\unit{cm}$;
\item $\left(\chi^{2}/\mathrm{dof}\right)_\mathrm{SV} < 5$, where dof is the number of degrees of freedom.
\end{itemize}
Muons are also required to satisfy additional quality criteria: 
the $\left(\chi^{2}/\mathrm{dof}\right)_\text{track}$ of each muon track is required to be less than 3, 
and the track itself is required to cross at least 6 layers (pixel or strip) of the tracker.

A large fraction of events contain only one OS muon pair and an associated SV.  
Among those events with more than one pair of OS muons, 
the pairs are ordered according to increasing $\chi^{2}_\mathrm{SV}$
and only the first or the first two pairs are used in this search.
A few of the selection criteria applied to the second muon pair are less stringent than those applied to the first pair 
in order to maximize the sensitivity to BSM signatures.

For the first (second) muon pair,
a muon is considered isolated if the track isolation, as defined in Section~\ref{sec:triggers}, is less than 10 (20)\% of \ptmu,
and if no jet is found within a cone of size $\Delta R<0.3$ around the muon itself.

As further described in Section~\ref{sec:strategy}, 
we explore isolated, partially isolated, and nonisolated dimuon topologies.
Isolated topologies consist of events where both muons are isolated;
partially isolated topologies consist of events where exactly one of the two muons fails the isolation criteria; 
finally, in nonisolated topologies both muons fail the isolation criteria.
In events with two muon pairs, the only category considered requires all four muons to be isolated.

For a BSM signal, we expect the dimuon system transverse momentum vector, \ptmumuvec, to be collinear with the SV vector, 
which is defined as the vector \vecdv connecting the PV to the SV. 
This does not apply to a muon pair where the two muons originate from different vertices
and erroneously form a common SV because of the accidental crossing of their trajectories.
In order to suppress such background, we require the azimuthal angle between the two vectors (\dphimudv) to be less than 0.02 (0.10)
for the first (second) muon pair.
Similarly, the opening angle between two muons erroneously forming a common SV tends to be very large, independently of the boost of the dimuon system.
Hence, in order to further reduce the background from events with the wrong muon SV association, we also require $\dphimumu<2.8$.
The main goal of such selections is to suppress background where SVs are formed from cosmic ray muons, muons from pileup collisions, and muons from back-to-back quantum chromodynamics dijet events.

Muons from pileup vertices can overlap in the $r$-$\phi$ plane despite being apart in $r$-$z$ at their origin, and,
as a consequence, they can be wrongly associated with a common SV.
These muon pairs are characterized by a large \detamumu and a small \dphimumu.
We exploit this feature to suppress this background contribution with the 
requirement $\logabsetaphi<1.25$.

Additional background sources include ``prompt'' (\ie, nondisplaced) muons and material interaction vertices, \ie, vertices from particles produced in interactions with detector material, \eg, in the beampipe or the pixel layers.
Muons from truly displaced vertices do not leave hits in the tracker layers between the interaction point and the SV.
In order to suppress the background from prompt muons, we therefore require that the observed number of hits found in the pixel tracker for each muon
be no larger than the number expected from the outward propagation of the muon itself from the associated SV.
This requirement is applied only when $\lxy>3.5\cm$ to ensure that the SV is located beyond the first pixel tracker layer.
Additionally, material interaction vertices
are suppressed by rejecting 
events with an SV located within 0.5\mm of a pixel tracker module. 

In order to identify displaced vertices of interest, we also exploit the muon track transverse impact parameter, \dxy.
To this end, we select events based on the \dxy significance ($\abs{\dxy}/\sigmadxy$) and the lifetime-scaled \dxy (\scaleddxy).
The first expression normalizes \dxy against its measurement uncertainty,
while the second normalizes \dxy against the lifetime corresponding to the observed SV transverse displacement. 
This allows the same selection requirements to be used for all lifetime hypotheses.
For the first (second) muon pair, we require each muon to satisfy: 
\begin{itemize}
\item $\abs{\dxy}/\sigmadxy > 2$ (1);
\item $\scaleddxy > 0.10$ (0.05).
\end{itemize}

The total selection efficiency 
after the trigger selections
typically ranges from about 55 to 75\% in simulated signal events, 
depending on the details of the models.

\subsection{Event categorization}
\label{sec:strategy}

This search targets long-lived low-mass dimuon resonances that are
characterized by displaced decays into pairs of muons.
Therefore, the primary feature to be exploited in order to discriminate a potential signal from the SM background
is the presence in each event of at least one pair of muons and an associated SV.
Most of the physics processes contributing to the SM background are
more highly suppressed at increased displacement of the SV from the PV.
Because the vertex position resolution along the beam ($z$ axis) is not as precise as the one achieved in the transverse ($xy$) plane,
we only use the \lxy of the SV to categorize dimuon events.
The \lxy categorization is intended to maximize the search sensitivity to a
range of potential BSM signal lifetimes, and is based on the 
CMS pixel tracker geometry.  We define the following six \lxy categories:

\begin{itemize} 
    \item within beam pipe: [0.0, 0.2]\cm; [0.2, 1.0]\cm; [1.0, 2.4]\cm;
    \item between beam pipe and first pixel tracker layer: [2.4, 3.1]\cm;
    \item between first and second pixel tracker layers: [3.1, 7.0]\cm;
    \item between second and third pixel tracker layers: [7.0, 11.0]\cm.
\end{itemize}

In each \lxy category, we further divide dimuon events in two bins of \ptmumu ($<25$ and $\geq25\GeV$). 
This categorization improves the sensitivity to different signal topologies since
\ptmumu tends to be low in signal models where long-lived particles arise from the decay of a \PQb hadron,
while it is expected to be larger in other models.
Finally, in each (\lxy, \ptmumu) category, dimuon events are  divided into bins of isolation (to distinguish isolated, partially isolated, and nonisolated topologies).

\begin{figure}[htb!]
  \centering
    \includegraphics[width=0.95\textwidth]{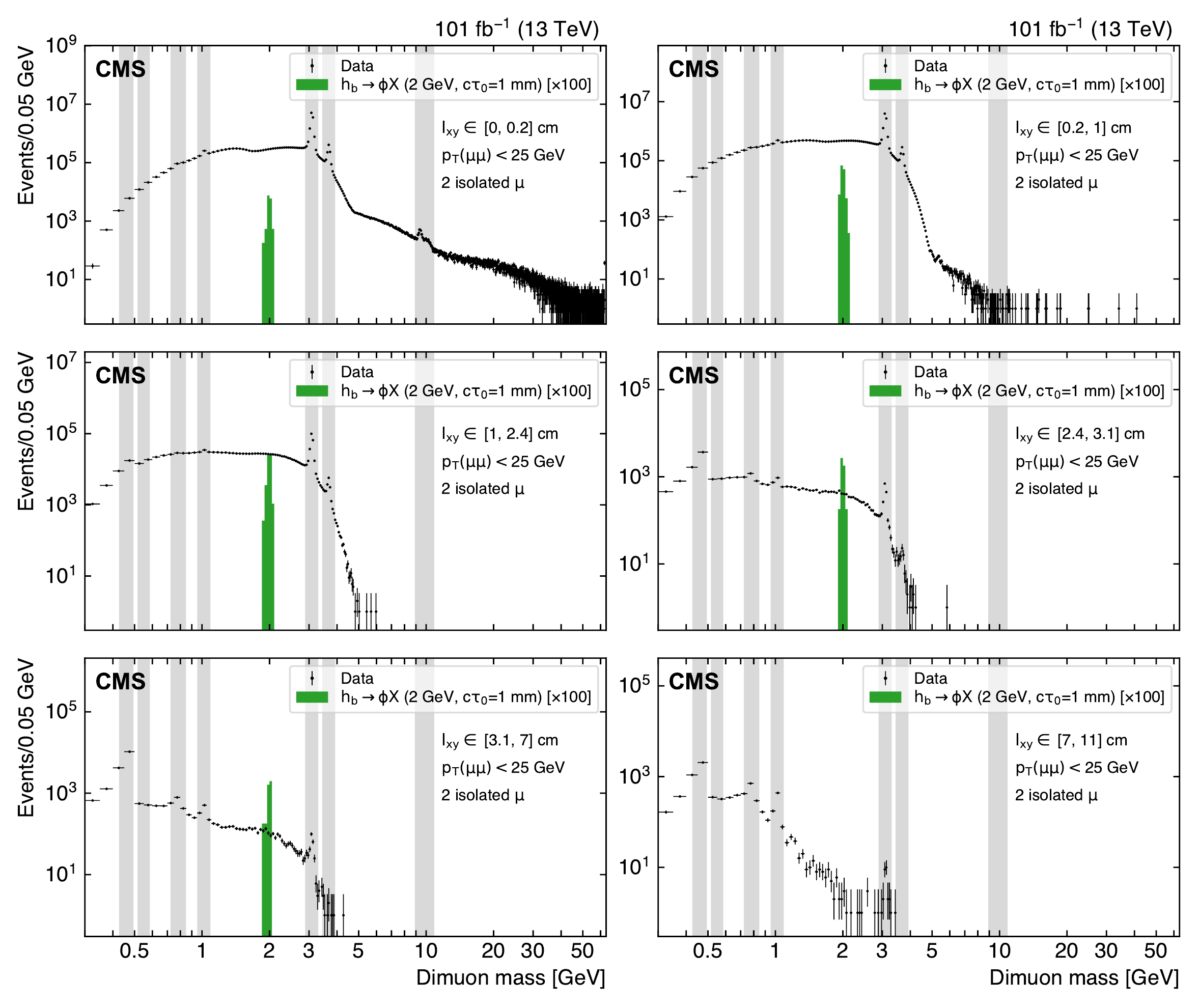}
    \caption{
      The dimuon invariant mass distribution is shown in bins of \lxy as obtained from selected dimuon events in data where both muons are isolated  
      with $\ptmumu<25\GeV$: (upper left) $0.0\leq\lxy<0.2\cm$; (upper right) $0.2\leq\lxy<1.0\cm$; (middle left) $1.0\leq\lxy<2.4\cm$; (middle right) $2.4\leq\lxy<3.1\cm$;
      (lower left) $3.1\leq\lxy<7.0\cm$; (lower right) $7.0\leq\lxy<11.0\cm$. 
      The distribution expected for a representative $h_{\PQb}\to\phi X$ signal model with $m_{\phi}=2\GeV$ and $c\tau_{0}^{\phi}=1\mm$ is overlaid.
      The signal event yield corresponds to a value of the branching fraction product
      $\mathcal{B}(h_{\PQb}\to\phi X) \, \mathcal{B}(\phi\to\mu\mu) = 1.2\times10^{-8}$, 
      equal to the median expected exclusion limit at 95\% \CL set in this paper (see Section~\ref{sec:results}),
      and is multiplied by a factor of 100 for display purposes.
      The vertical gray bands indicate mass ranges containing known SM resonances, which are masked for the purpose of this search.
      \label{fig:mass_pt0}} 
\end{figure}

\begin{figure}[htb!]
  \centering
    \includegraphics[width=0.95\textwidth]{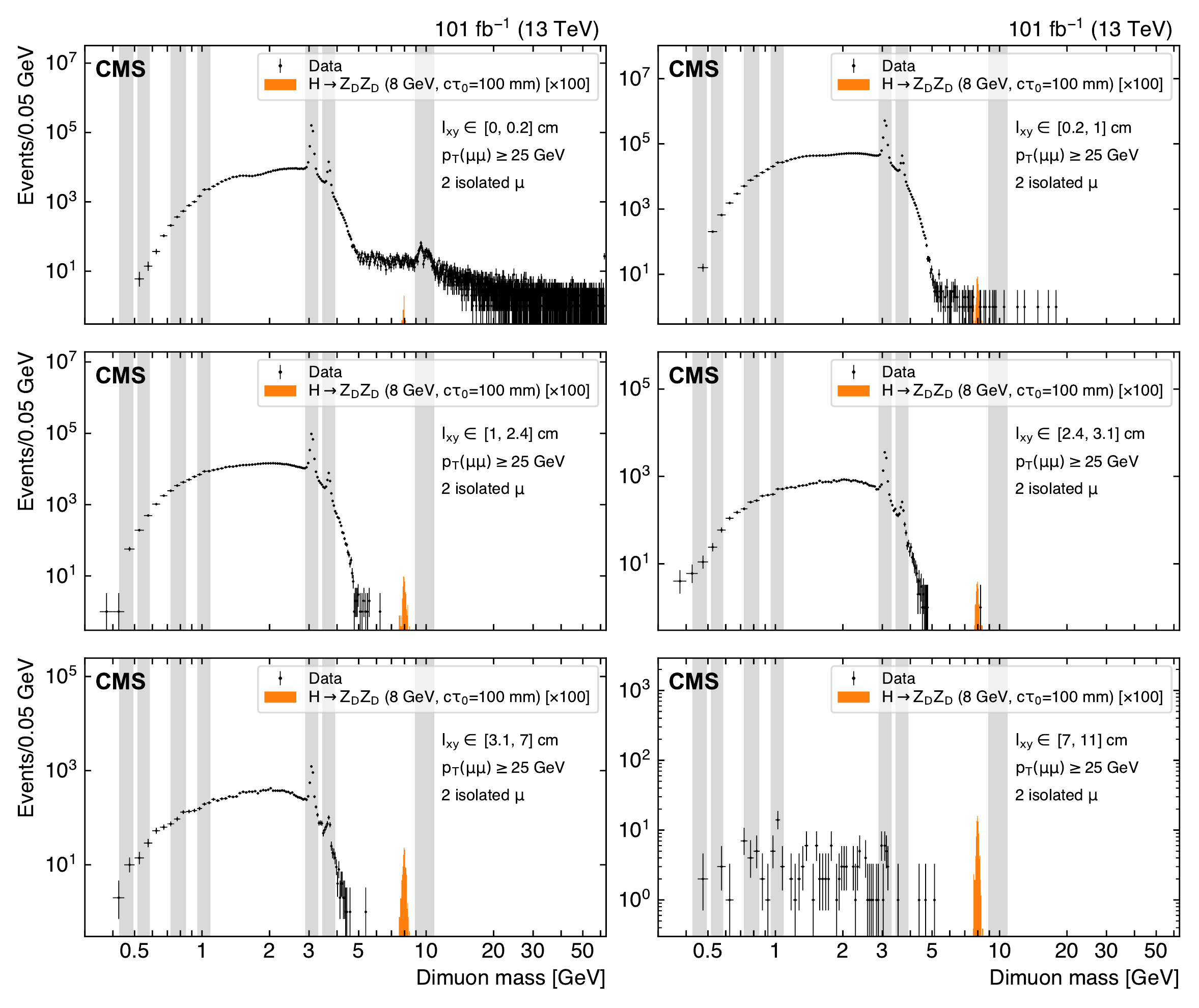}
    \caption{
      The dimuon invariant mass distribution is shown in bins of \lxy as obtained from selected dimuon events in data 
      where both muons are isolated with $\ptmumu\geq25\GeV$: 
      (upper left) $0.0\leq\lxy<0.2\cm$; (upper right) $0.2\leq\lxy<1.0\cm$; 
      (middle left) $1.0\leq\lxy<2.4\cm$; (middle right) $2.4\leq\lxy<3.1\cm$;
      (lower left) $3.1\leq\lxy<7.0\cm$; (lower right) $7.0\leq\lxy<11.0\cm$. 
      The distribution expected for a representative $\PH\to\PZ_{\mathrm{D}}\PZ_{\mathrm{D}}$ signal model 
      with $m_{\PZ_{\mathrm{D}}}=8\GeV$ and $c\tau_{0}^{\PZ_{\mathrm{D}}}=100\mm$, 
      corresponding to $\epsilon=1.2 \times 10^{-7}$, is overlaid.
      The signal event yield corresponds to a value of the branching fraction product 
      $\mathcal{B}(\PH\to\PZ_{\mathrm{D}}\PZ_{\mathrm{D}}) \, \mathcal{B}(\PZ_{\mathrm{D}}\to\PGm\PGm) = 1.2\times10^{-5}$, 
      equal to the median expected exclusion limit at 95\% \CL set in this paper (see Section~\ref{sec:results}),
      and is multiplied by a factor of 100 for display purposes.
      The vertical gray bands indicate mass ranges containing known SM resonances, which are masked for the purpose of this search.
      \label{fig:mass_pt1}}  
\end{figure}

In each (\lxy, \ptmumu, isolation) bin, we define mass windows sliding along the dimuon invariant mass spectrum (see Section~\ref{sec:fit}),
and we perform a search for a resonant dimuon peak in each mass window. 
Mass regions corresponding to
known resonances decaying either in dimuon pairs or in a pair of charged hadrons are not considered, i.e., they are ``masked''.
As an example, in Fig.~\ref{fig:mass_pt0} (\ref{fig:mass_pt1}) we show the \mmumu distribution in bins of \lxy
as obtained from selected dimuon events where both muons are isolated and have \ptmumu below (above) 25\GeV.

In events with two muon pairs each associated with an SV, 
we further require the difference between the two dimuon masses to be within 5\% of their mean,
and the four-muon mass to be consistent with the mass of the SM Higgs boson ($115 < m_{4\PGm} < 135\GeV$).
The selected four-muon events are treated as an exclusive independent category,
aimed at improving the search sensitivity to models of BSM physics where an SM Higgs boson decays to a pair of $\PZ_{\mathrm{D}}$ bosons, each decaying to two muons.
After applying all selection criteria, we observe zero events in this category.

In each dimuon mass window, the SM background is modeled by fitting the dimuon invariant mass distribution in data, 
as further described in Section~\ref{sec:fit}.
Finally, the dimuon mass windows in all (\lxy, \ptmumu, isolation) bins and the single four-muon event category are used in combination
as input to a final likelihood fit for the interpretation of the results, as described in Section~\ref{sec:results}.
In the four-muon category, the two dimuon invariant mass distributions are not fitted separately.

\section{Parametrization of the invariant mass distribution}
\label{sec:fit}

The search is performed by fitting mass distributions in the various categories to the sum of signal plus
SM background models.  
We target low-mass dimuon resonances whose intrinsic width is assumed to be much narrower than the detector resolution.
The signal dimuon mass resolution ($\sigma_{\text{mass}}^{\PGm\PGm}$) is estimated from simulated events to be about 1.1\% of the mass, 
with an uncertainty of about 50\%, and both are independent of the signal mass and lifetime hypotheses.
The signal four-muon mass resolution is similarly estimated from simulated $\PH\to\PZ_{\mathrm{D}}\PZ_{\mathrm{D}}\to 4\PGm$ events
to be about 1.1\% of the Higgs boson mass, with an uncertainty of about 50\%.
We use the sum of a Gaussian function and a double-sided Crystal Ball function~\cite{dcb1,dcb2} to model signal dimuon (four-muon) invariant mass distributions in dimuon (four-muon) events.

In each dimuon event category, 
the SM background is modeled in sliding windows of ${\pm}5\sigma_{\text{mass}}^{\PGm\PGm}$ around the considered signal mass hypothesis.
These windows are required to not overlap with the masked mass regions corresponding to known SM resonances (see Section~\ref{sec:strategy}).
Instead, in the four-muon event category we consider a single window centered around the known Higgs boson mass ($115 < m_{4\PGm} < 135\GeV$).
A number of functional forms are considered in order to model the background mass distribution.  These include
Bernstein polynomials, exponential functions, as well as combinations of the two.  
For each family of functional forms,
the best order in each mass window and event category is selected 
by means of a Fisher test~\cite{ftest}.
First, the lowest order ($N$) function is used to fit the data.
Then, the next-order ($N+1$) is used, and the difference
$2\Delta\mathrm{NLL}_{N+1}=2\left(\mathrm{NLL}_{N+1}-\mathrm{NLL}_{N}\right)$ 
(with NLL denoting the negative logarithm of the likelihood of the fit)
is evaluated to determine whether the data support the need for a higher order
function. 
This decision is based on the fact that 
$2\Delta\mathrm{NLL}_{N+1}$ is asymptotically described by a $\chi^{2}$ distribution with $M$ degrees of freedom, 
where $M$ is the difference in the number of free parameters in the $(N+1)$th and $N$th order functions. 
A $p$-value is calculated as $P_M\left(\chi^{2} > 2\Delta\mathrm{NLL}_{N+1}\right)$ where $P_M\left(\chi_{\text{min}}^2\right)$ 
is the $\chi^{2}$ tail probability for $M$ degrees of freedom.
If the $p$-value is less than 0.05, 
the higher order function is retained, 
since it is determined to significantly improve the description of the data.
Once the best order $N$ for each family of functions has been determined, 
the corresponding functional forms are entered in an envelope then used 
in the \textit{discrete profiling} method~\cite{discreteprofile}, 
where the choice of the background function is treated as a discrete nuisance parameter in the fit
to account for the uncertainty associated with the arbitrary choice of the function.
For Bernstein polynomials, we also include orders $N-1$ (if the selected best order $N>0$) and $N+1$ in the list of suitable functions.
Typically, the selected best order $N$ is $\leq4$ and is smaller in event categories with a small number of observed events.
In addition, we use a 
goodness-of-fit test to remove background models that do not describe the data appropriately.
The test is based on a $\chi^{2}$ test statistic, which is converted into a $p$-value
through a set of pseudo-experiments.  Models with $p < 0.01$ are not considered.
The potential presence of a bias in the measurement of a signal due to the choice of the functional forms used to model the SM background 
is also assessed by means of pseudo-experiments. 
First, a varying amount of signal is injected on top of the background generated according to the selected functional form. 
Then, a background+signal fit is performed allowing the signal yield to float freely. 
The bias, quantified as the difference between the measured and injected signal yields relative to the statistical uncertainty in the measured signal yield, 
is found to be negligible.

\begin{figure}[htb!]
  \centering
    \includegraphics[width=0.85\textwidth]{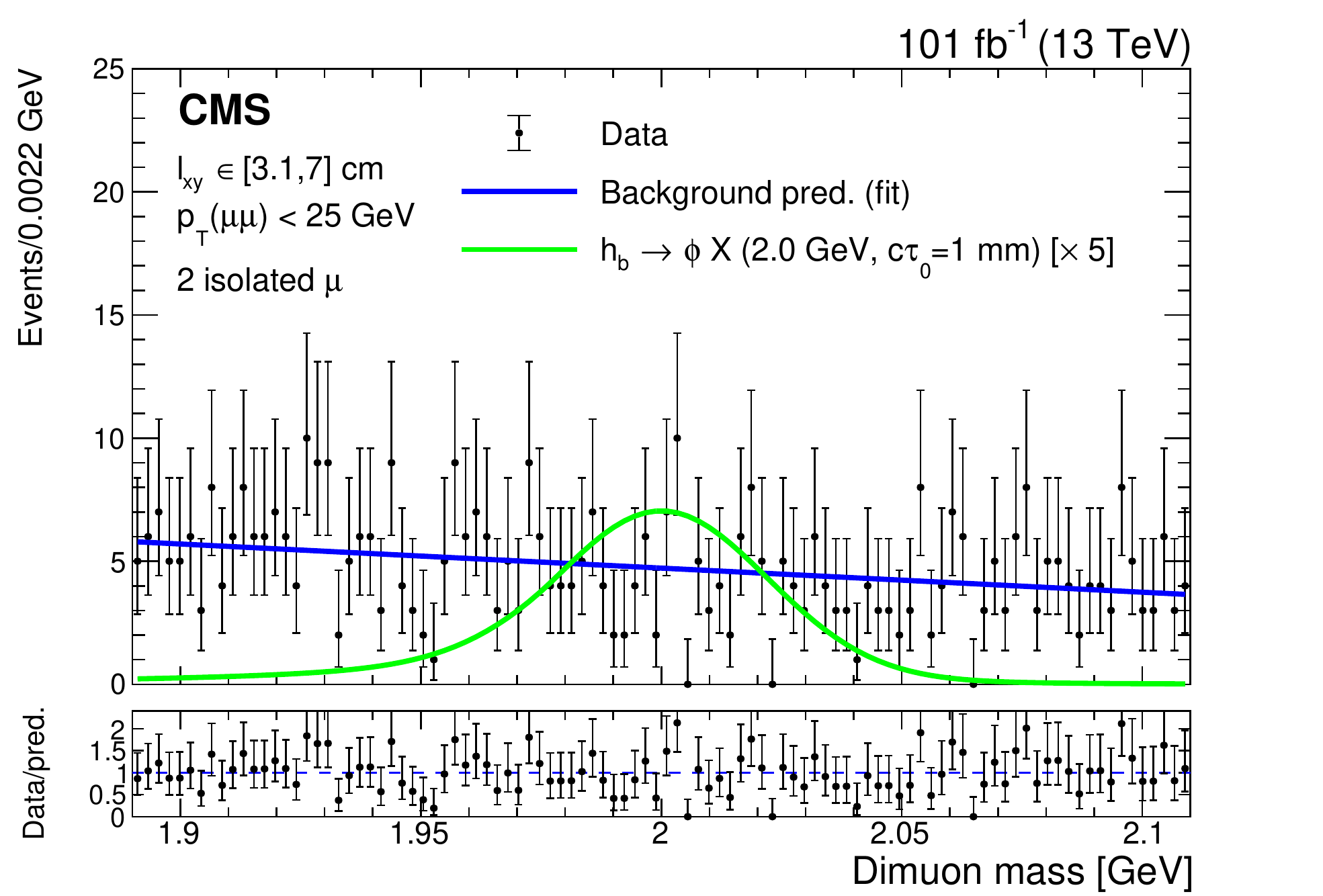}
    \caption{
      The dimuon invariant mass distribution is shown in a mass window around 2\GeV, 
      in one of the dimuon search bins ($3.1\leq\lxy<7.0\cm$, $\ptmumu<25\GeV$, with two isolated muons).
      The result of the background-only fit to the data is also shown (blue)
      together with the dimuon invariant mass distribution expected for a representative $h_{\PQb}\to\phi X$ signal model (green)
      with $m_{\phi}=2\GeV$ and $c\tau_{0}^{\phi}=1\mm$.
      The signal event yield corresponds to a value of the branching fraction product
      $\mathcal{B}(h_{\PQb}\to\phi X) \, \mathcal{B}(\phi\to\mu\mu) = 1.2\times10^{-8}$, 
      equal to the median expected exclusion limit at 95\% \CL set in this paper (see Section~\ref{sec:results}),
      and is multiplied by a factor of 5 for display purposes.
      The lower panel shows the ratio of the data to the background prediction, with error bars corresponding 
      to the statistical uncertainty in the data.
      \label{fig:fit}}
\end{figure}

The background-only fit results for the selected functional forms in one of the dimuon search bins are shown in Fig.~\ref{fig:fit}
together with the dimuon invariant mass distribution expected for a representative signal.

\section{Results and interpretation}
\label{sec:results}

Binned maximum likelihood fits to the data
are performed simultaneously in all search bins, 
under either background-only or background+signal hypotheses, using background and signal
models and uncertainties described in Section~\ref{sec:fit}.
Additional log-normal constraint terms are used to account 
for the uncertainties in the signal yields, when considered (Section~\ref{sec:signalsysts}).

No significant peak-like structures are observed in data.
The background+signal fits are used to set 95\% confidence level (\CL) upper limits 
on the branching fractions for the signal models under consideration.  
Limits are set using a modified frequentist approach, employing the \CLs 
criterion~\cite{Read:2002hq,Junk:1999kv,Cowan:2010js,ATLAS:2011tau}.
These limits are then used, in conjunction
with the theoretical cross section calculations, 
to exclude ranges of masses and/or lifetimes 
for the BSM particles of the signal models.

\subsection{Systematic uncertainties in the signal yields}
\label{sec:signalsysts}

A systematic uncertainty in the trigger efficiency
(see Section~\ref{sec:triggers}) is assessed 
that ranges between 3 and 30\%, 
depending on the \pt and displacement of the muons.
It is derived from studies of data collected with different triggers 
and accounts for differences between the measurement and the simulation. 
In order to account for potential mismodeling of the signal simulation,
we also assess a systematic uncertainty in the signal yields
equal to 100\% of the inefficiency of the selection (see Section~\ref{sec:evtsel})
applied after the trigger selections.
The resulting systematic uncertainty in the signal yield is then 
between 25 and 45\%, depending on the signal mass and lifetime.
Additionally, we assess a systematic uncertainty in the expected signal yields
arising from the uncertainty in the luminosity measurement of 2.3\%~\cite{LUM-17-004} for 2017 and 2.5\%~\cite{LUM-18-002} for
2018 data.
Finally, we account for uncertainties resulting from the limited sizes of the simulated signal samples.
Uncertainties arising from the choice of the PDF and 
of the renormalization and factorization 
scales used in the event generator are negligible 
compared to others.
Uncertainties in the trigger efficiency and integrated luminosity are treated 
as correlated across search bins. Other uncertainties are taken as uncorrelated.
Since we find no significant difference between 2017 and 2018 data and simulations,
all uncertainties are treated as correlated across data taking periods.

\subsection{Constraints on models of BSM physics}

\begin{figure}[htb!]
  \centering
    \includegraphics[width=0.85\textwidth]{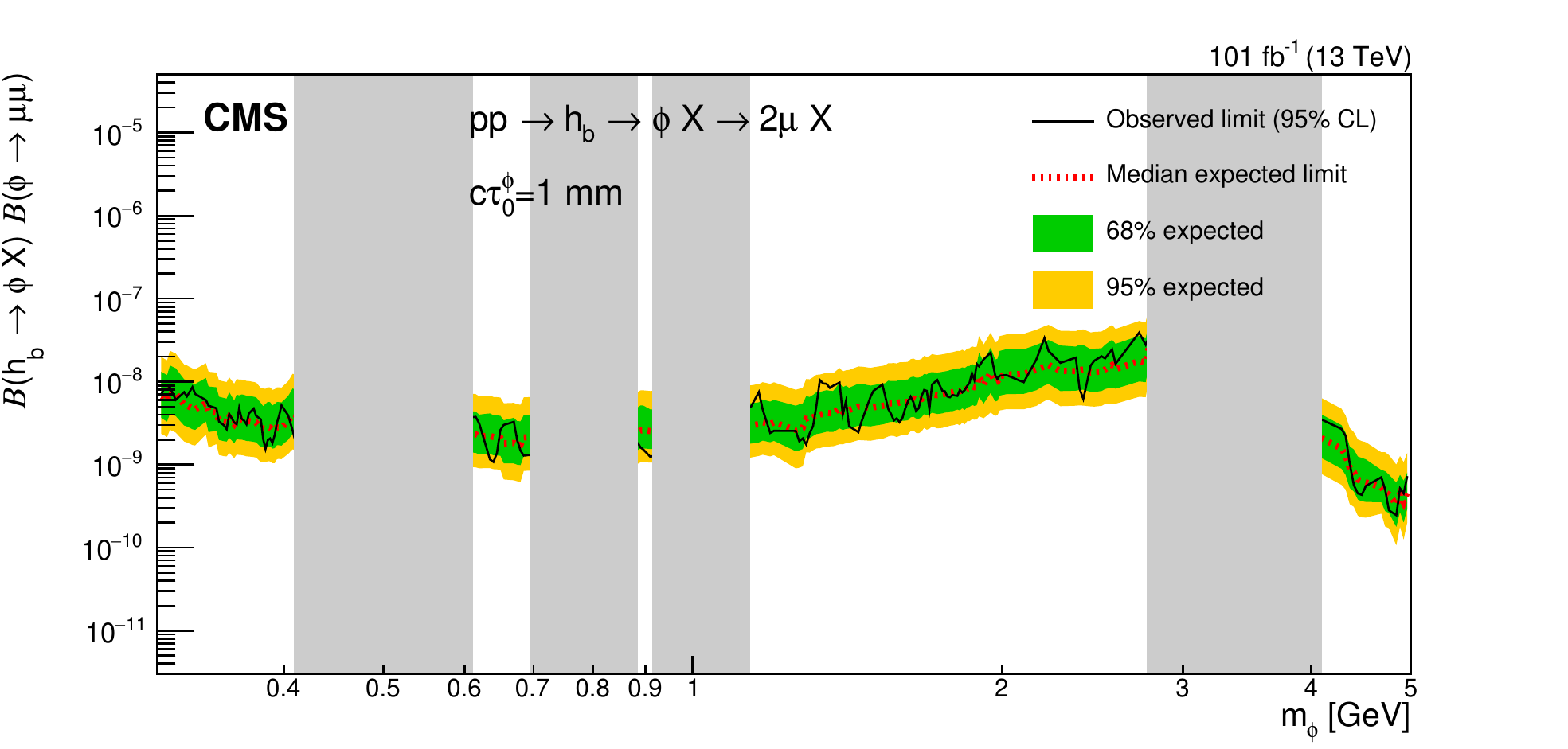}
    \includegraphics[width=0.85\textwidth]{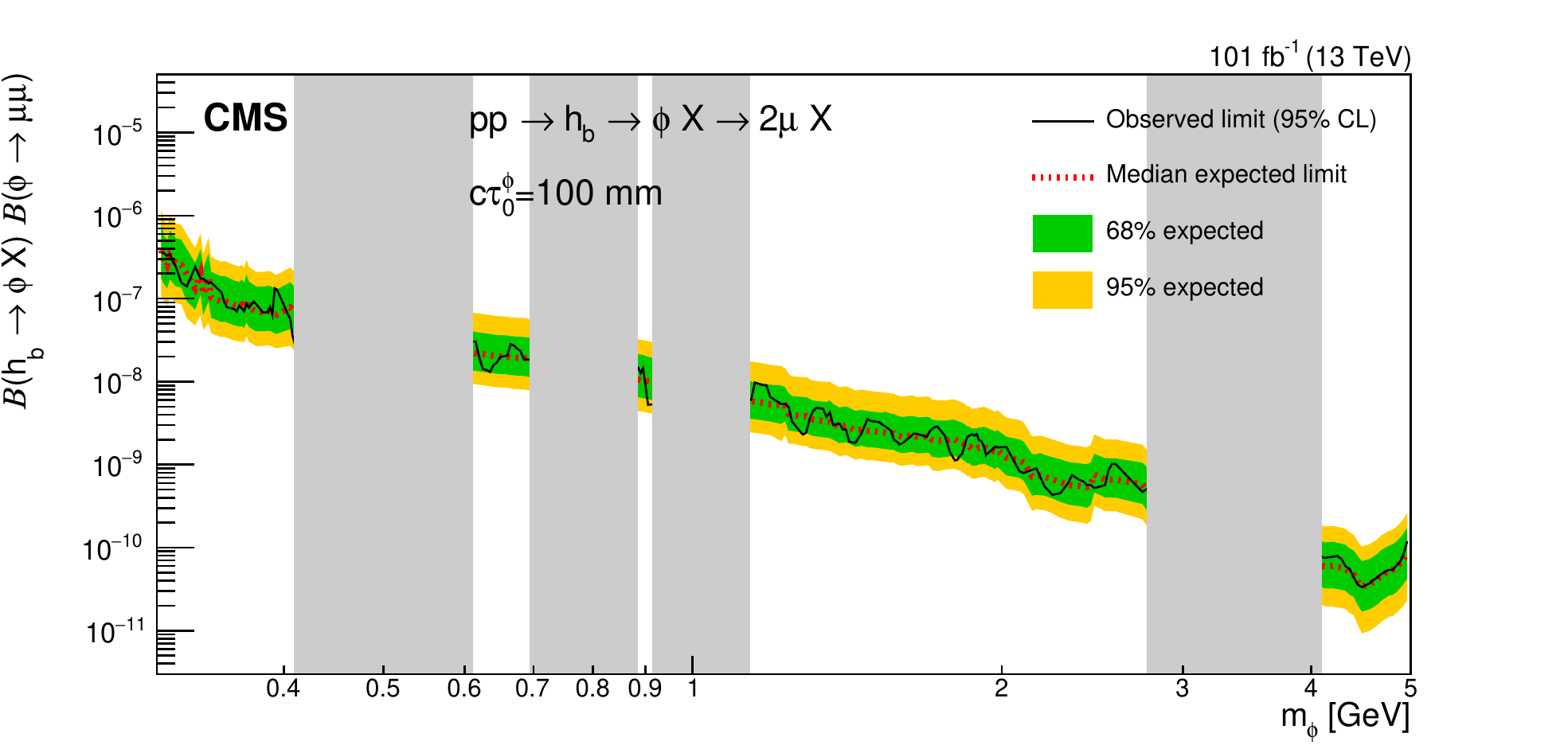}
    \caption{
      Exclusion limits at 95\% \CL 
      on the branching fraction product $\mathcal{B}(h_{\PQb}\to\phi X) \, \mathcal{B}(\phi\to\mu\mu)$,
      as functions of the signal mass ($m_{\phi}$)
      for $c\tau_{0}^{\phi}=1\mm$ (upper)
      and 100\mm (lower).
      The solid black (dashed red) line represents the observed (median expected) exclusion. 
      The inner green (outer yellow) band indicates the region containing 68 (95)\% of the distribution of limits expected 
      under the background-only hypothesis.
      The vertical gray bands indicate mass ranges containing known SM resonances,
      which are masked for the purpose of this search. 
      The limits are obtained using the combination of all dimuon event categories.
      \label{fig:bphi}}
\end{figure}

\begin{figure}[htb!]
  \centering
    \includegraphics[width=0.85\textwidth]{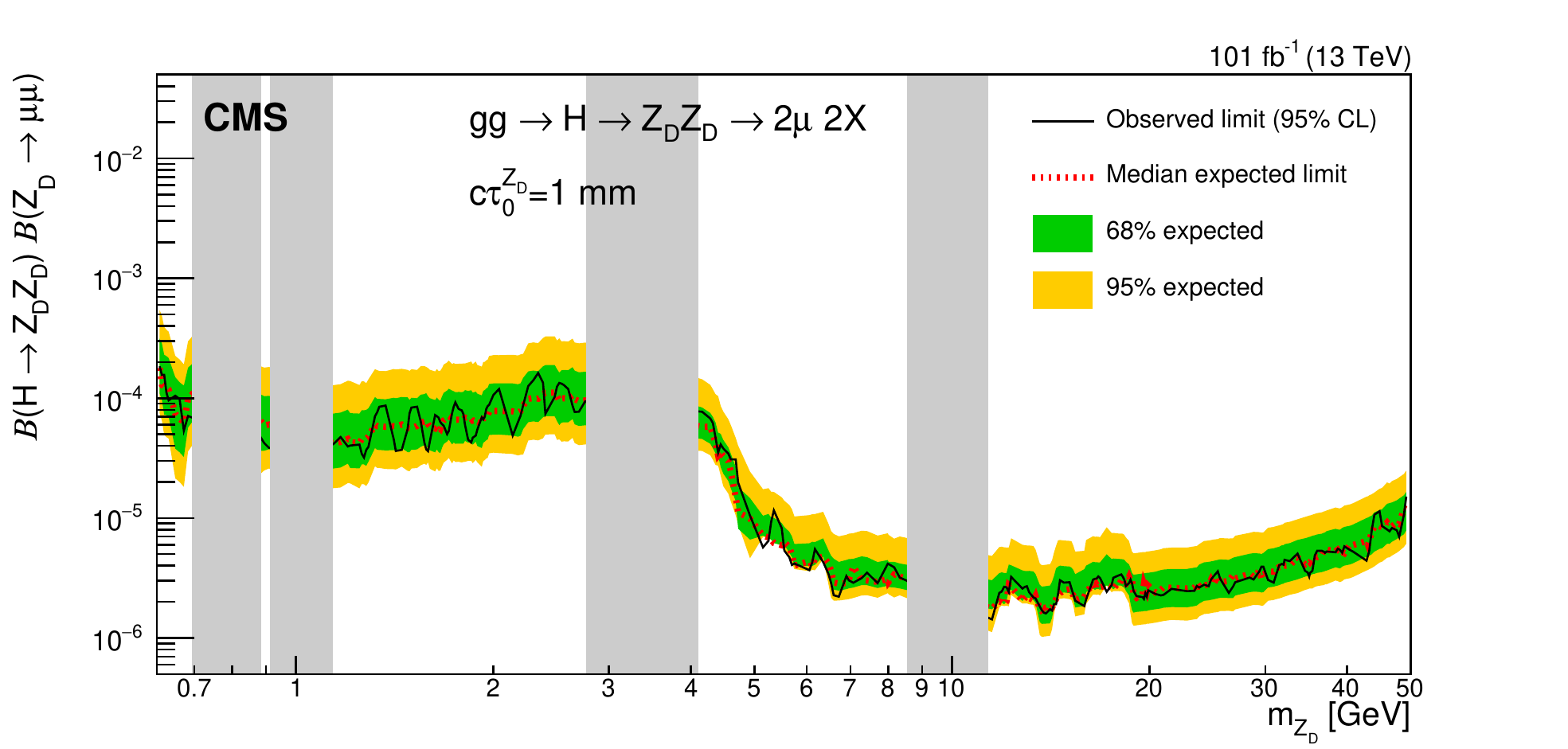}
    \includegraphics[width=0.85\textwidth]{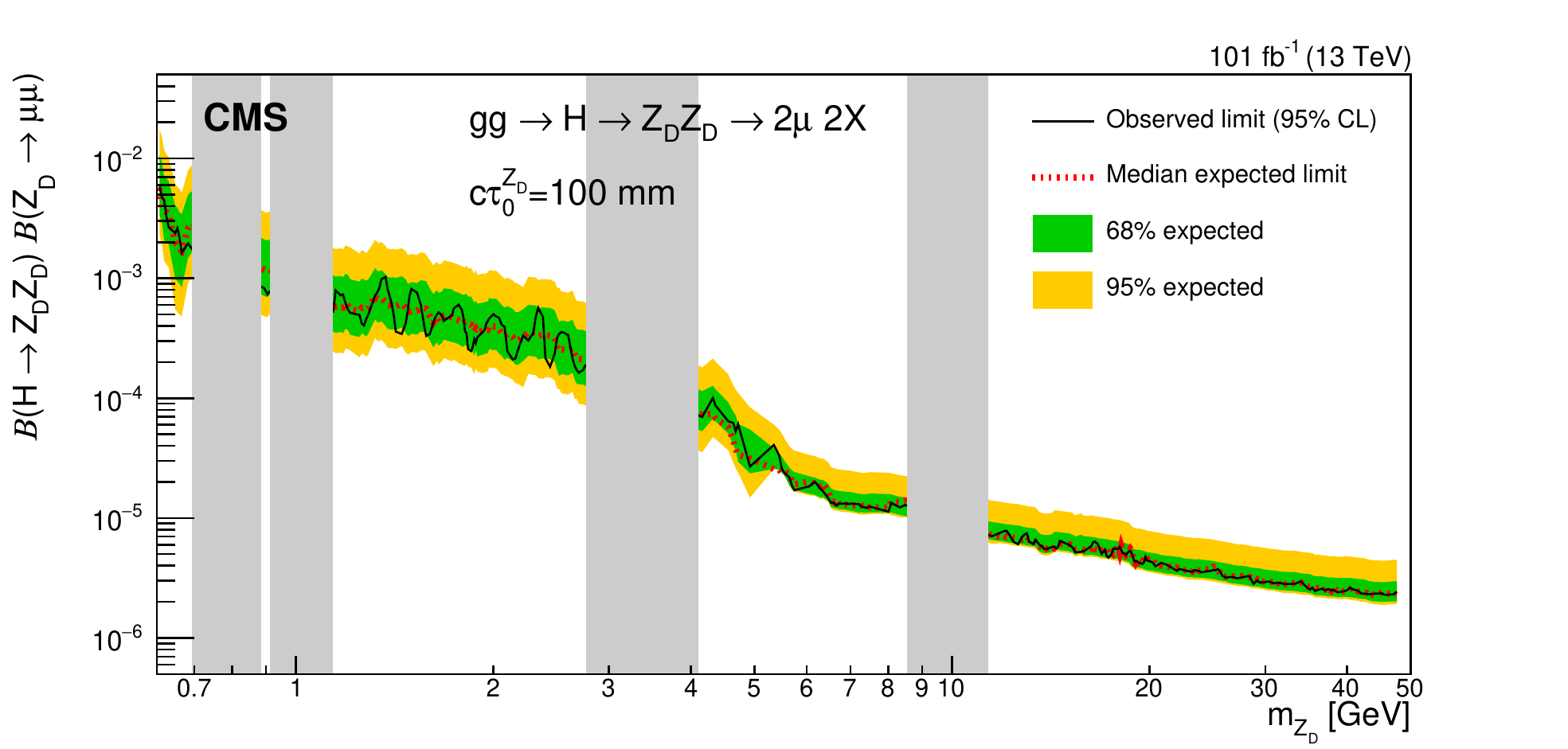}
    \caption{
      Exclusion limits at 95\% \CL 
      on the branching fraction product $\mathcal{B}(\PH\to\PZ_{\mathrm{D}}\PZ_{\mathrm{D}}) \, \mathcal{B}(\PZ_{\mathrm{D}}\to\PGm\PGm)$,
      as functions of the signal mass ($m_{\PZ_{\mathrm{D}}}$) 
      for $c\tau_{0}^{\PZ_{\mathrm{D}}}=1\mm$ (upper)
      and 100\mm (lower).
      The solid black (dashed red) line represents the observed (median expected) exclusion. 
      The inner green (outer yellow) band indicates the region containing 68 (95)\% of the distribution of limits expected 
      under the background-only hypothesis.
      At $m_{\PZ_{\mathrm{D}}}\gtrsim5\GeV$ the bands are asymmetrical,
      due to the small number of events observed in the corresponding search bins, especially those with $\lxy>0.2\cm$.
      The vertical gray bands indicate mass ranges containing known SM resonances,
      which are masked for the purpose of this search. 
      The limits are obtained using the combination of all dimuon event categories.
      \label{fig:hzd}}  
\end{figure}

\begin{figure}[htb!]
  \centering
    \includegraphics[width=0.85\textwidth]{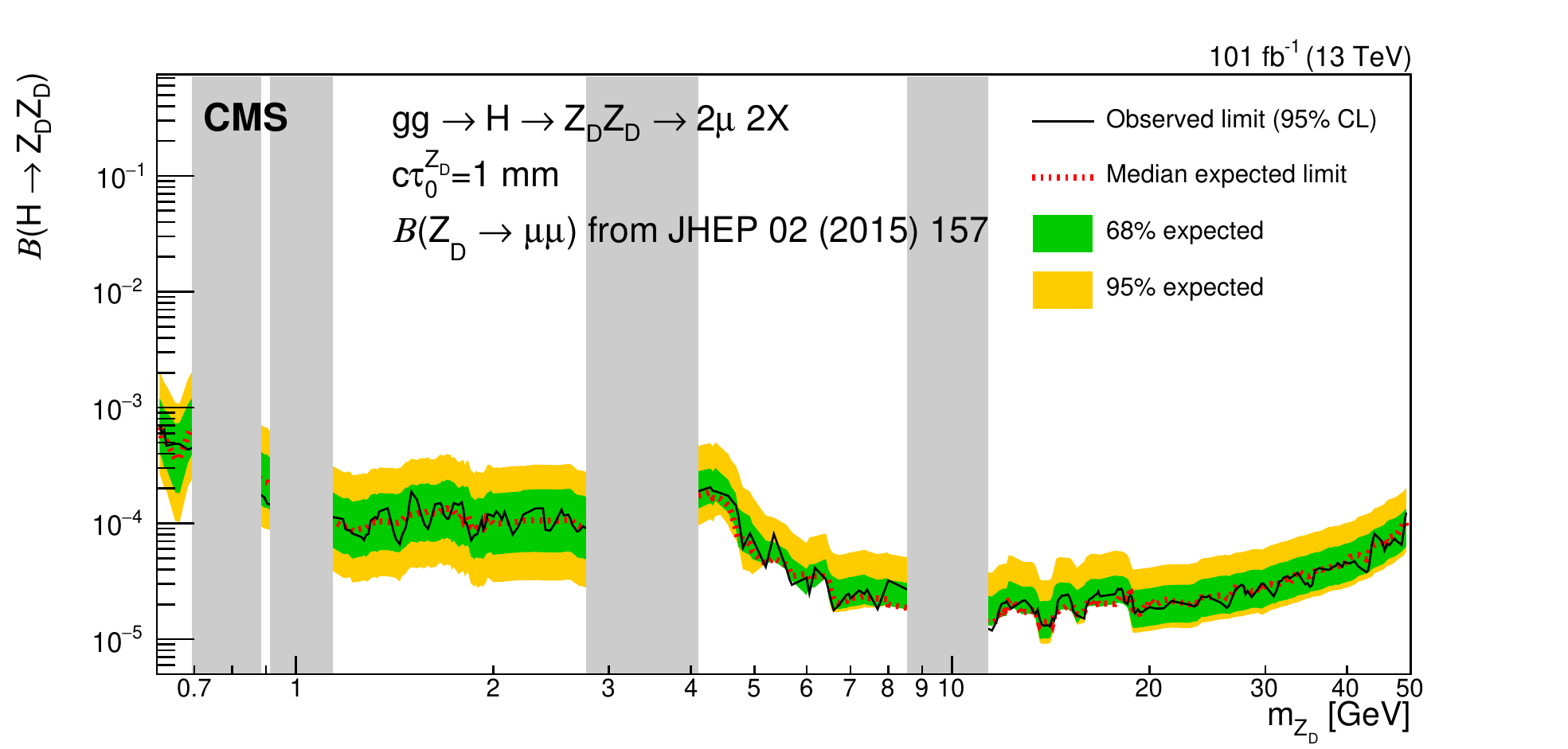}
    \includegraphics[width=0.85\textwidth]{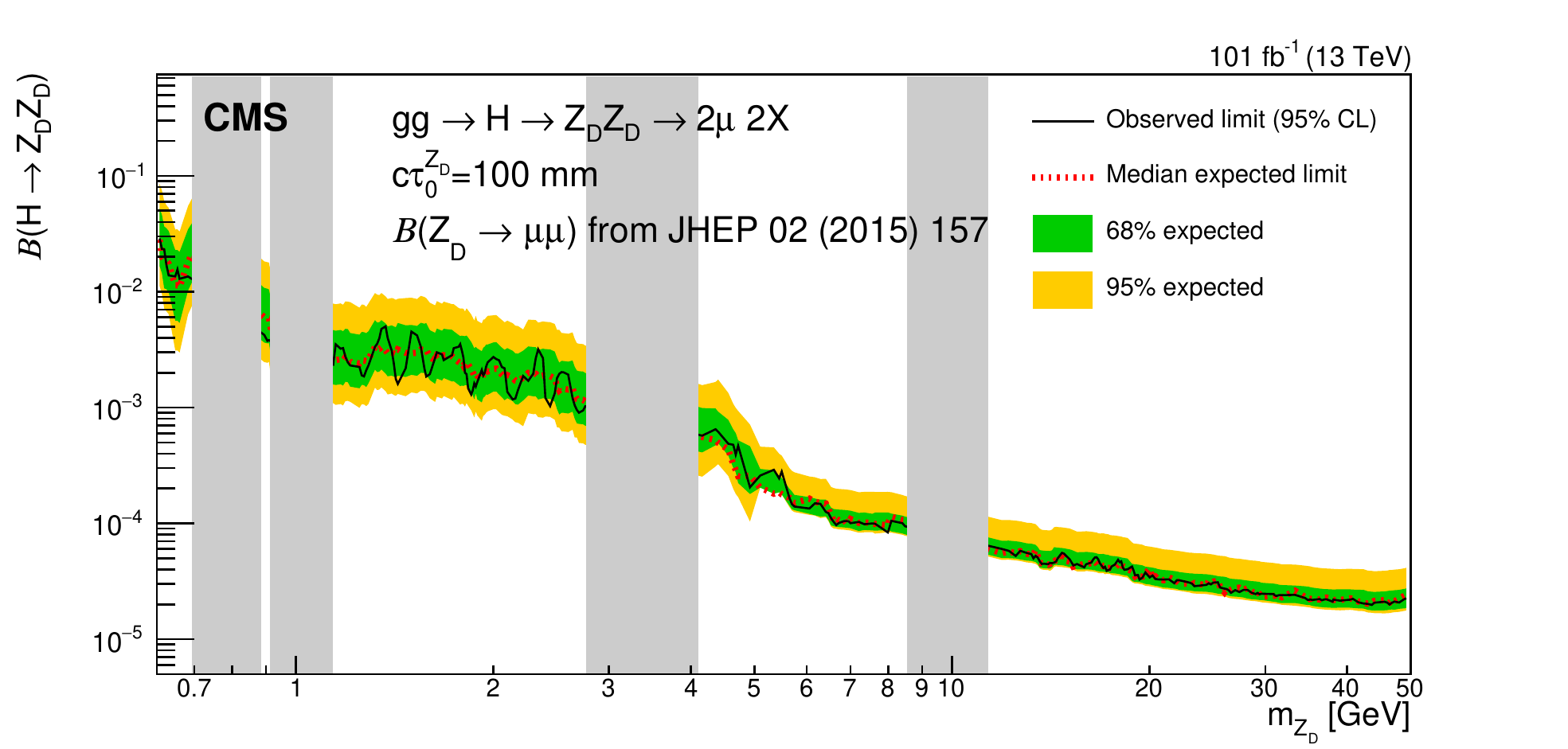}
    \caption{
      Exclusion limits at 95\% \CL 
      on the branching fraction $\mathcal{B}(\PH\to\PZ_{\mathrm{D}}\PZ_{\mathrm{D}})$,
      as functions of the signal mass ($m_{\PZ_{\mathrm{D}}}$) 
      for $c\tau_{0}^{\PZ_{\mathrm{D}}}=1\mm$ (upper)
      and 100\mm (lower),
      for the $\PH\to\PZ_{\mathrm{D}}\PZ_{\mathrm{D}}$ signal model, assuming
      values of $\mathcal{B}(\PZ_{\mathrm{D}}\to\PGm\PGm)$ from the model of Ref.~\cite{DM6}.
      The solid black (dashed red) line represents the observed (median expected) exclusion. 
      The inner green (outer yellow) band indicates the region containing 68 (95)\% of the distribution of limits expected 
      under the background-only hypothesis.
      At $m_{\PZ_{\mathrm{D}}}\gtrsim5\GeV$ the bands are asymmetrical,
      due to the small number of events observed in the corresponding search bins, especially those with $\lxy>0.2\cm$.
      The vertical gray bands indicate mass ranges containing known SM resonances,
      which are masked for the purpose of this search. 
      The limits are obtained using the combination of all dimuon and four-muon event categories.
      \label{fig:hzdall}} 
\end{figure}

We set upper limits at 95\% \CL on the branching fractions
for the $h_{\PQb}\to\phi X$ and $\PH\to\PZ_{\mathrm{D}}\PZ_{\mathrm{D}}$ 
signal models.
For the $h_{\PQb}\to\phi X$ model, Fig.~\ref{fig:bphi}
shows the upper limits on the branching fraction product 
$\mathcal{B}(h_{\PQb}\to\phi X) \, \mathcal{B}(\phi\to\mu\mu)$, 
as a function of the signal mass hypothesis,
for a few representative lifetime hypotheses.
For the $\PH\to\PZ_{\mathrm{D}}\PZ_{\mathrm{D}}$ model,
the upper limits on the branching fraction product 
$\mathcal{B}(\PH\to\PZ_{\mathrm{D}}\PZ_{\mathrm{D}}) \, \mathcal{B}(\PZ_{\mathrm{D}}\to\PGm\PGm)$,
as a function of the signal mass, are shown in Fig.~\ref{fig:hzd}.
In addition, Fig.~\ref{fig:hzdall} shows the upper limits on the branching fraction 
$\mathcal{B}(\PH\to\PZ_{\mathrm{D}}\PZ_{\mathrm{D}})$,
using values of $\mathcal{B}(\PZ_{\mathrm{D}}\to\PGm\PGm)$ from the model of Ref.~\cite{DM6}.
The upper limits shown in Figs.~\ref{fig:bphi} and \ref{fig:hzd} are obtained using only the combination of dimuon event categories.
For the limits shown in Fig.~\ref{fig:hzdall}, the four-muon event category is added to the combination, 
since in this case the relative acceptances for signal events in the dimuon and four-muon categories 
can be determined using the $\mathcal{B}(\PZ_{\mathrm{D}}\to\PGm\PGm)$ assumption.

Constraints for signals with mass above the \PQb hadron mass are more stringent since the background from displaced muons from $h_{\PQb}$ decays is completely eliminated.
The constraints tend to get weaker at longer lifetime because of the loss of acceptance due to decays beyond the scouting trigger acceptance,
which is determined by the radial coverage of the pixel detector.  
We note that our limits on the model with a long-lived scalar emitted in $h_{\PQb}$ decays cannot be directly compared
to similar limits from the LHCb experiment~\cite{BtoPhiLimits-1,BtoPhiLimits-2}, since our limits are on the
inclusive $h_{\PQb}\to\phi X$ branching fraction, while the LHCb limits are on the exclusive 
$\PBz\to\phi \PKst(892)^{0}$ and $\PBp\to\phi \PKp$ branching fractions. However, we scale our inclusive limits based on the fraction of \PBz 
or \PBp production from MC and thus compare our exclusive branching fraction limits to those of LHCb. Such comparisons are provided as supplemental material~\cite{HEPData}.

\begin{figure}[htb!]
  \centering
    \includegraphics[width=0.85\textwidth]{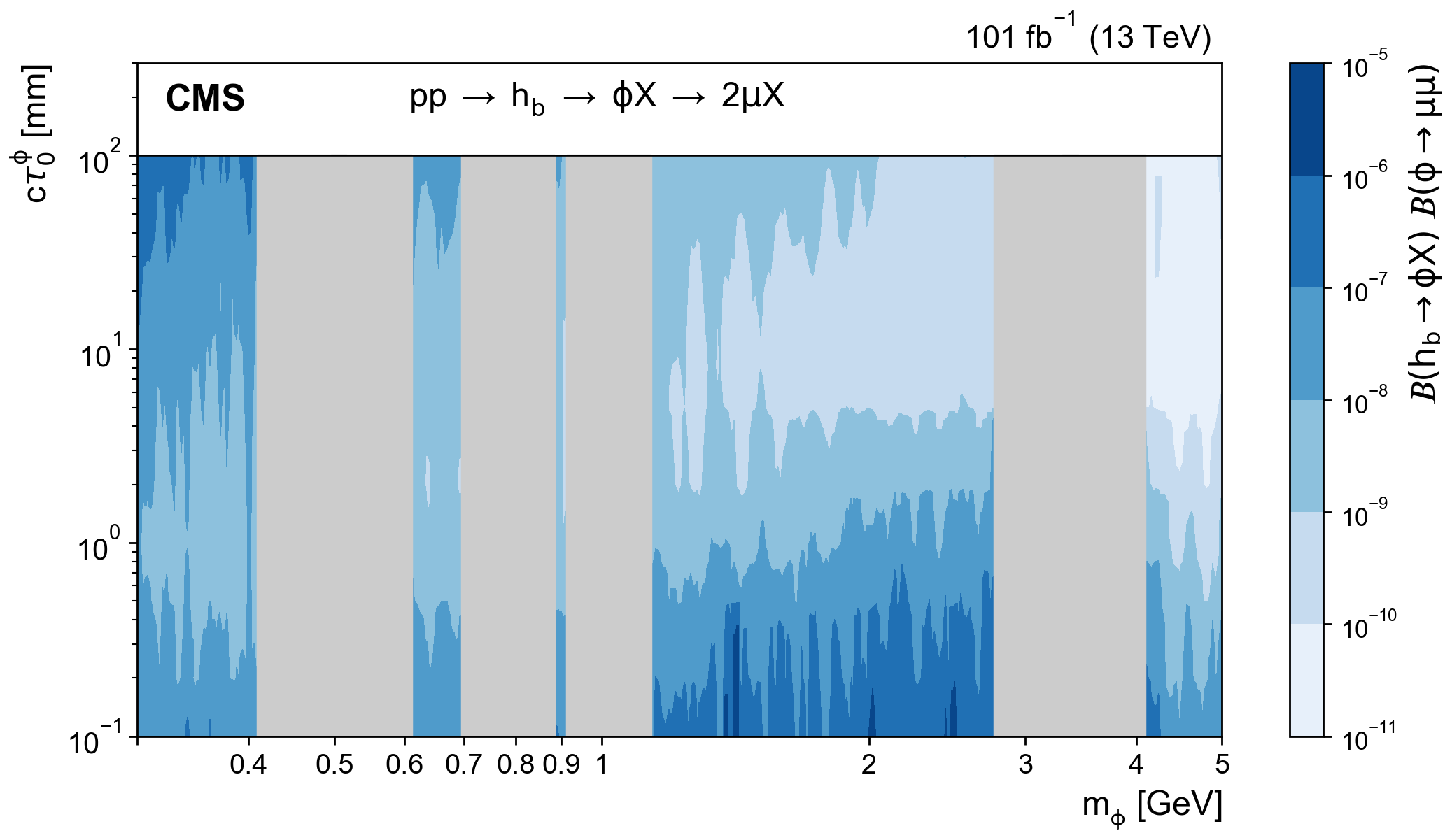}
    \caption{
      Observed limits at 95\% CL on the branching fraction product $\mathcal{B}(h_{\PQb}\to\phi X) \, \mathcal{B}(\phi\to\mu\mu)$ as contours in the parameter space containing the signal mass ($m_{\phi}$) and the signal lifetime $c\tau_{0}^{\phi}$.
      The vertical gray bands indicate mass ranges containing known SM resonances, which are masked for the purpose of this search.
      The limits are obtained using the combination of all dimuon event categories.
    \label{fig:bphicontours}}  
\end{figure}

\begin{figure}[htb!]
  \centering
    \includegraphics[width=0.85\textwidth]{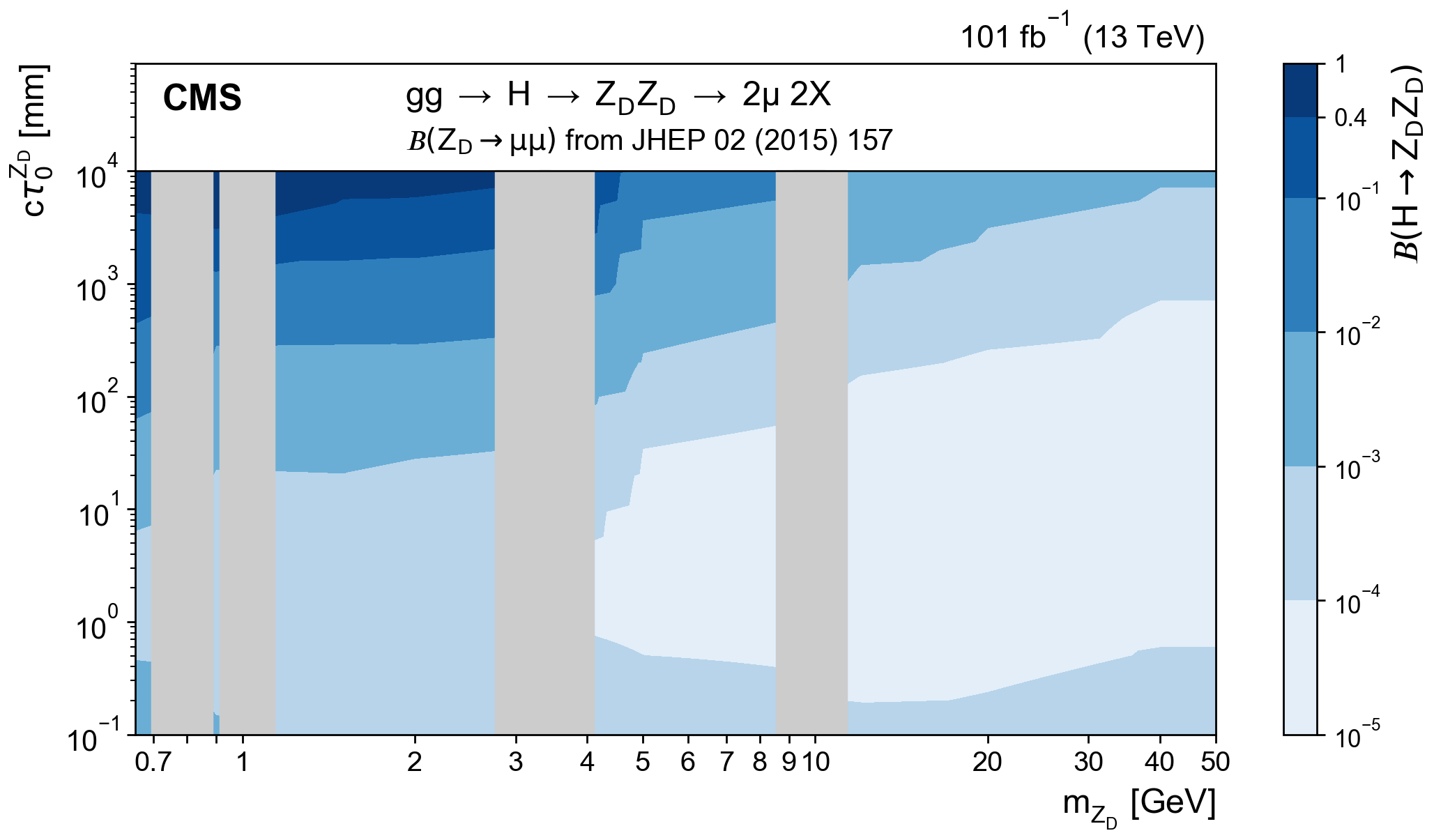}
    \caption{
      Observed limits at 95\% CL on the branching fraction $\mathcal{B}(\PH\to\PZ_{\mathrm{D}}\PZ_{\mathrm{D}})$ as contours in the parameter space containing the signal mass ($m_{\PZ_{\mathrm{D}}}$) and the signal lifetime $c\tau_{0}^{\PZ_{\mathrm{D}}}$ for the $\PH\to\PZ_{\mathrm{D}}\PZ_{\mathrm{D}}$ signal model assuming
      values of $\mathcal{B}(\PZ_{\mathrm{D}}\to\PGm\PGm)$ from the model of Ref.~\cite{DM6}.
      The vertical gray bands indicate mass ranges containing known SM resonances, which are masked for the purpose of this search.
      The limits are obtained using the combination of all dimuon and four-muon event categories.
    \label{fig:hzdallcontours}}  
\end{figure}

Observed limits at 95\% CL as contours in the parameter space of signal mass and signal lifetime for both the models considered in this search are shown in Fig.~\ref{fig:bphicontours} and Fig.~\ref{fig:hzdallcontours}.

\begin{figure}[htb!]
  \centering
    \includegraphics[width=0.85\textwidth]{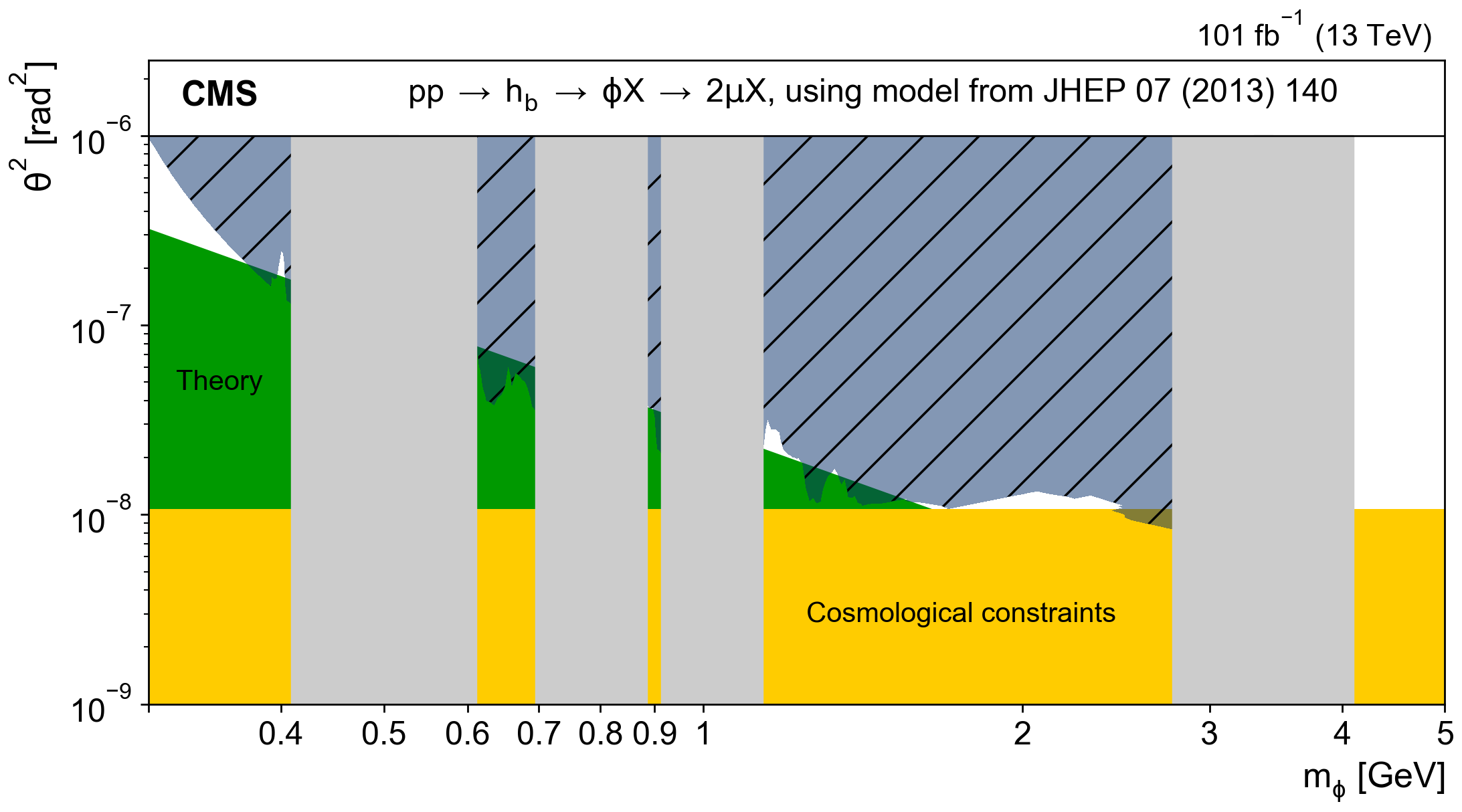}
    \caption{
      Region excluded at 95\% CL using this search (shown in hatched gray) for the inflaton model~\cite{BPhi-inflaton}
      in the parameter space containing the signal mass ($m_{\phi}$) and the square of the signal mixing angle ($\theta^{2}$). The regions forbidden by theory and cosmological constraints are shown in green and yellow respectively. The vertical gray bands indicate mass ranges containing known SM resonances, which are masked for the purpose of this search.
      The limits are obtained using the combination of all dimuon event categories.
    \label{fig:theta}} 
\end{figure}

\begin{figure}[htb!]
  \centering
    \includegraphics[width=0.85\textwidth]{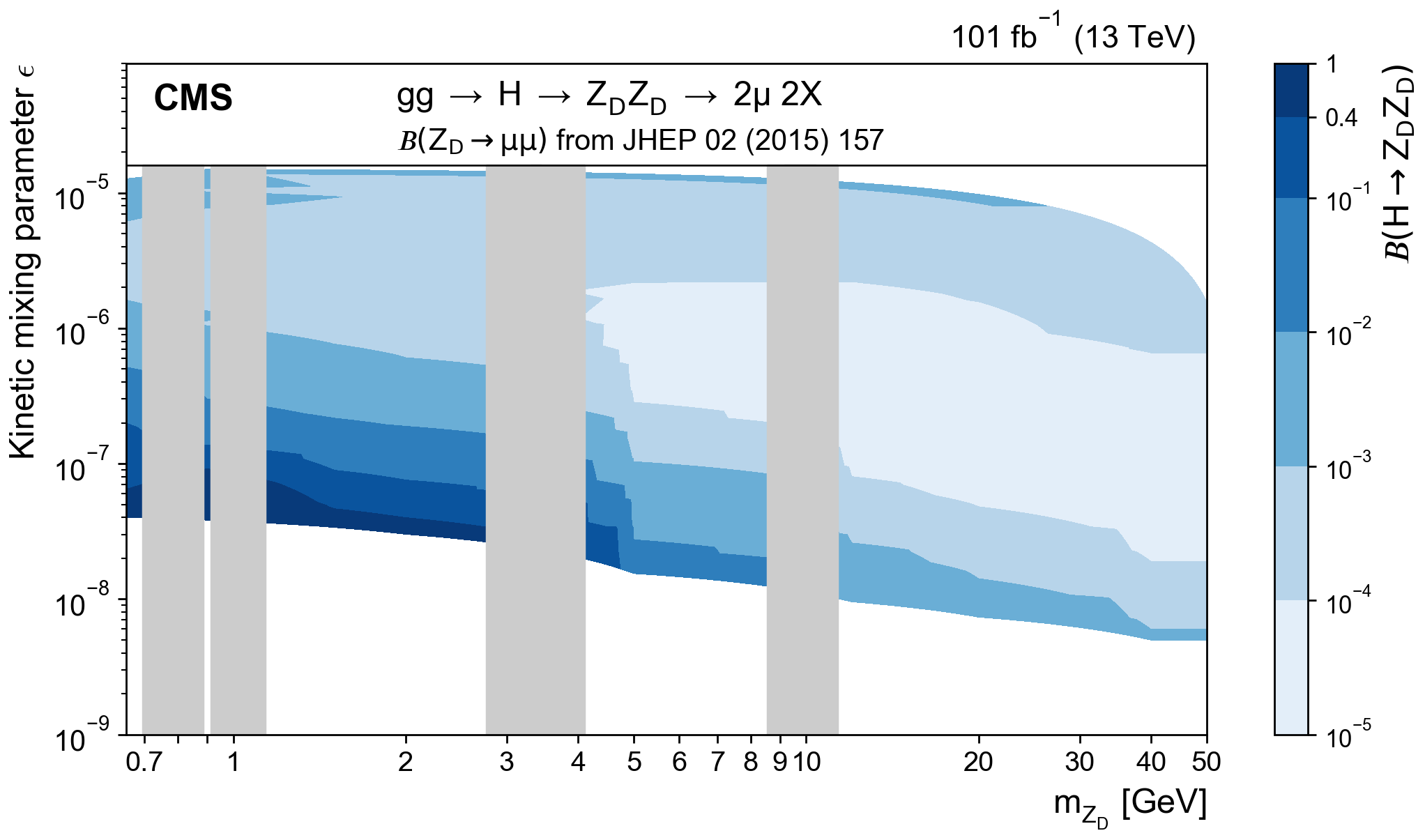}
    \caption{
      Observed limits at 95\% CL on the branching fraction $\mathcal{B}(\PH\to\PZ_{\mathrm{D}}\PZ_{\mathrm{D}})$ as contours in the parameter space containing the signal mass ($m_{\PZ_{\mathrm{D}}}$) and the signal kinetic mixing parameter $\epsilon$ for the $\PH\to\PZ_{\mathrm{D}}\PZ_{\mathrm{D}}$ signal model
      assuming values of $\mathcal{B}(\PZ_{\mathrm{D}}\to\PGm\PGm)$ from the model of Ref.~\cite{DM6}.
      The vertical gray bands indicate mass ranges containing known SM resonances, which are masked for the purpose of this search.
      The white regions at the top (bottom) correspond to values of $c\tau_0^{\PZ_{\mathrm{D}}}<0.1\mm$ ($>10^{4}\mm$).
      The limits are obtained using the combination of all dimuon and four-muon event categories.
    \label{fig:epsilon}}  
\end{figure}

In Fig.~\ref{fig:theta}, we show the excluded region at 95\% CL for the inflaton model~\cite{BPhi-inflaton}
in the parameter space containing the signal mass ($m_{\phi}$) and the square of the signal mixing angle ($\theta^{2}$) using 
the results from the $h_{\PQb}\to\phi X$ signal model. 
The signal mixing angle $\theta$ determines the inflaton's coupling to the SM fields via mixing with the Higgs boson. In Fig.~\ref{fig:epsilon}, we show the observed limits at 95\% CL as contours in the parameter space containing the signal mass ($m_{\PZ_{\mathrm{D}}}$) and the signal kinetic mixing coupling $\epsilon$ for the $\PH\to\PZ_{\mathrm{D}}\PZ_{\mathrm{D}}$ signal model.

\clearpage

Tabulated results are provided in the HEPData record for this analysis~\cite{HEPData}.
To facilitate the reinterpretation of these results in other BSM models, we also provide upper limits at 95\% \CL on the number of observed events 
as a function of the dimuon mass in a number of signal regions characterized by different \lxy ranges, with and without an isolation requirement, inclusive in \ptmumu as well as with $\ptmumu\geq25\GeV$, together with parametrizations of the event selection efficiency.

\section{Summary}
\label{ref:summary}

A search for displaced dimuon resonances has been performed using proton-proton collisions at a center-of-mass energy of 13\TeV, 
collected by the CMS experiment at the LHC in 2017 and 2018, corresponding to an integrated luminosity of \lumivalscouting.
The data sets used in this search are collected using a dedicated dimuon trigger stream with low transverse momentum thresholds, 
recorded at high rate by retaining a reduced amount of information, 
in order to explore otherwise inaccessible phase space at low dimuon mass and nonzero displacement from the primary interaction vertex.
No significant excess beyond the standard model expectation is found,
and the data are used to set constraints on a wide range of mass and lifetime hypotheses for models of physics beyond the standard model 
where a Higgs boson decays to a pair of long-lived dark photons, or where a long-lived scalar resonance arises from the decay of a \PQb hadron.
These constraints are the most stringent to date for substantial regions of the parameter space.

\begin{acknowledgments}
  We congratulate our colleagues in the CERN accelerator departments for the excellent performance of the LHC and thank the technical and administrative staffs at CERN and at other CMS institutes for their contributions to the success of the CMS effort. In addition, we gratefully acknowledge the computing centers and personnel of the Worldwide LHC Computing Grid and other centers for delivering so effectively the computing infrastructure essential to our analyses. Finally, we acknowledge the enduring support for the construction and operation of the LHC, the CMS detector, and the supporting computing infrastructure provided by the following funding agencies: BMBWF and FWF (Austria); FNRS and FWO (Belgium); CNPq, CAPES, FAPERJ, FAPERGS, and FAPESP (Brazil); MES and BNSF (Bulgaria); CERN; CAS, MoST, and NSFC (China); MINCIENCIAS (Colombia); MSES and CSF (Croatia); RIF (Cyprus); SENESCYT (Ecuador); MoER, ERC PUT and ERDF (Estonia); Academy of Finland, MEC, and HIP (Finland); CEA and CNRS/IN2P3 (France); BMBF, DFG, and HGF (Germany); GSRI (Greece); NKFIA (Hungary); DAE and DST (India); IPM (Iran); SFI (Ireland); INFN (Italy); MSIP and NRF (Republic of Korea); MES (Latvia); LAS (Lithuania); MOE and UM (Malaysia); BUAP, CINVESTAV, CONACYT, LNS, SEP, and UASLP-FAI (Mexico); MOS (Montenegro); MBIE (New Zealand); PAEC (Pakistan); MSHE and NSC (Poland); FCT (Portugal); JINR (Dubna); MON, RosAtom, RAS, RFBR, and NRC KI (Russia); MESTD (Serbia); MCIN/AEI and PCTI (Spain); MOSTR (Sri Lanka); Swiss Funding Agencies (Switzerland); MST (Taipei); ThEPCenter, IPST, STAR, and NSTDA (Thailand); TUBITAK and TAEK (Turkey); NASU (Ukraine); STFC (United Kingdom); DOE and NSF (USA).

 \hyphenation{Rachada-pisek} Individuals have received support from the Marie-Curie program and the European Research Council and Horizon 2020 Grant, contract Nos.\ 675440, 724704, 752730, 758316, 765710, 824093, 884104, and COST Action CA16108 (European Union); the Leventis Foundation; the Alfred P.\ Sloan Foundation; the Alexander von Humboldt Foundation; the Belgian Federal Science Policy Office; the Fonds pour la Formation \`a la Recherche dans l'Industrie et dans l'Agriculture (FRIA-Belgium); the Agentschap voor Innovatie door Wetenschap en Technologie (IWT-Belgium); the F.R.S.-FNRS and FWO (Belgium) under the ``Excellence of Science -- EOS" -- be.h project n.\ 30820817; the Beijing Municipal Science \& Technology Commission, No. Z191100007219010; the Ministry of Education, Youth and Sports (MEYS) of the Czech Republic; the Deutsche Forschungsgemeinschaft (DFG), under Germany's Excellence Strategy -- EXC 2121 ``Quantum Universe" -- 390833306, and under project number 400140256 - GRK2497; the Lend\"ulet (``Momentum") Program and the J\'anos Bolyai Research Scholarship of the Hungarian Academy of Sciences, the New National Excellence Program \'UNKP, the NKFIA research grants 123842, 123959, 124845, 124850, 125105, 128713, 128786, and 129058 (Hungary); the Council of Science and Industrial Research, India; the Latvian Council of Science; the Ministry of Science and Higher Education and the National Science Center, contracts Opus 2014/15/B/ST2/03998 and 2015/19/B/ST2/02861 (Poland); the Funda\c{c}\~ao para a Ci\^encia e a Tecnologia, grant CEECIND/01334/2018 (Portugal); the National Priorities Research Program by Qatar National Research Fund; the Ministry of Science and Higher Education, projects no. 14.W03.31.0026 and no. FSWW-2020-0008, and the Russian Foundation for Basic Research, project No.19-42-703014 (Russia); MCIN/AEI/10.13039/501100011033, ERDF ``a way of making Europe", and the Programa Estatal de Fomento de la Investigaci{\'o}n Cient{\'i}fica y T{\'e}cnica de Excelencia Mar\'{\i}a de Maeztu, grant MDM-2017-0765 and Programa Severo Ochoa del Principado de Asturias (Spain); the Stavros Niarchos Foundation (Greece); the Rachadapisek Sompot Fund for Postdoctoral Fellowship, Chulalongkorn University and the Chulalongkorn Academic into Its 2nd Century Project Advancement Project (Thailand); the Kavli Foundation; the Nvidia Corporation; the SuperMicro Corporation; the Welch Foundation, contract C-1845; and the Weston Havens Foundation (USA).

\end{acknowledgments}

\bibliography{auto_generated}
\cleardoublepage \appendix\section{The CMS Collaboration \label{app:collab}}\begin{sloppypar}\hyphenpenalty=5000\widowpenalty=500\clubpenalty=5000\cmsinstitute{Yerevan~Physics~Institute, Yerevan, Armenia}
A.~Tumasyan
\cmsinstitute{Institut~f\"{u}r~Hochenergiephysik, Vienna, Austria}
W.~Adam\cmsorcid{0000-0001-9099-4341}, J.W.~Andrejkovic, T.~Bergauer\cmsorcid{0000-0002-5786-0293}, S.~Chatterjee\cmsorcid{0000-0003-2660-0349}, M.~Dragicevic\cmsorcid{0000-0003-1967-6783}, A.~Escalante~Del~Valle\cmsorcid{0000-0002-9702-6359}, R.~Fr\"{u}hwirth\cmsAuthorMark{1}, M.~Jeitler\cmsAuthorMark{1}\cmsorcid{0000-0002-5141-9560}, N.~Krammer, L.~Lechner\cmsorcid{0000-0002-3065-1141}, D.~Liko, I.~Mikulec, P.~Paulitsch, F.M.~Pitters, J.~Schieck\cmsAuthorMark{1}\cmsorcid{0000-0002-1058-8093}, R.~Sch\"{o}fbeck\cmsorcid{0000-0002-2332-8784}, D.~Schwarz, S.~Templ\cmsorcid{0000-0003-3137-5692}, W.~Waltenberger\cmsorcid{0000-0002-6215-7228}, C.-E.~Wulz\cmsAuthorMark{1}\cmsorcid{0000-0001-9226-5812}
% ...
\vskip\cmsinstskip
% ...
1:~Also at TU Wien, Wien, Austria\\
% ...
\end{sloppypar}
\end{document}